\documentclass[aps,prd,twocolumn,superscriptaddress,showpacs]{revtex4-2}
\usepackage{graphicx}
\usepackage{amsmath}
\usepackage{amsfonts}
\usepackage{amssymb}
\usepackage{amssymb}
\usepackage{slashed}
\usepackage{pstricks,pst-coil}
\usepackage{dsfont}
\usepackage{bm}
\usepackage{footnote}

\begin{document}
\title{The Lee-Wick-Chern-Simons pseudo-quantum electrodynamics}
%
%


\author{M. J. Neves}\email{mariojr@ufrrj.br}

\affiliation{Departamento de F\'isica, Universidade Federal Rural do Rio de Janeiro, BR 465-07, 23890-971, Serop\'edica, Rio de Janeiro, Brazil}
\begin{abstract}
The Lee-Wick pseudo-quantum electrodynamics in the presence of a Chern-Simons term is studied in this paper. 
The paper starts with a non-local lagrangian density that sets the pseudo-Lee-Wick electrodynamics defined on a 
$1+2$ space-time added to a non-local Chern-Simons topological term. 
Thus, we obtain the Lee-Wick-Chern-Simons pseudo-electrodynamics as a most complete gauge invariant model 
that provides a light mass associated with the Chern-Simons parameter, and also includes a Lee-Wick heavy mass. We investigate classical aspects as the potential 
energy for the interaction of static charges through the gauge propagator. The causality of theory is discussed through the retarded 
Green function in the coordinate space. The gauge field of the Lee-Wick-Chern-Simons pseudo-electrodynamics 
is minimally coupled to the fermions sector that includes new degree of freedoms, as a Lee-Wick heavy fermion partner of the electron.
The perturbative approach for the theory is presented via effective action in which we obtain the Ward identities. We study the quantum 
corrections at one loop, as the electron self-energy, the vacuum polarization, and the $3$-vertex.  
We show that the Lee-Wick mass has a fundamental role in these results, where it works like a natural regulator of the ultraviolet divergences. 
The $g-2$ factor for the electron is obtained as function of the LW mass, and of the CS parameter.  
Through the optical theorem, the Lee-Wick-Chern-Simons pseudo-electrodynamics is unitary at the tree level.

\end{abstract}
\maketitle

%
%



\section{Introduction}

%
The connection between quantum field theories (QFTs) in planar spaces and condensed matter physics (CMP) 
has called attention for the research in material physics. Several applications of field theories in low dimensions 
has been used to explain the experimental results in materials, as the transport in graphene 
\cite{Gorbar,Herbut,Gusynin,Herbut2}, Dirac materials \cite{Castro}, the quantum Hall effect 
\cite{Ando,Tsui,Laughlin,Chamon}, topological insulators \cite{Qi,Hasan,Chiu,Zhao,Qi2008}, superconductivity in layered materials \cite{Tesanovic,Zhang,Franz,Kivelson,Marino2018}, and others.
Abelian gauge field theory is the theoretical description for the mediators of the interaction between electron-positrons in planar materials, 
similar to usual quantum electrodynamics (QED). The condition that constraints the classical sources to be confined to a bidimensional space 
leads to the non-local gauge field theory, that is known in the literature as pseudo-electrodynamics (PED) \cite{Marino93}. The non-locality manifests
through the derivatives of infinity order in the D'Alembertian operator. The PED preserves the fundamental properties of a QFT 
as causality \cite{Amaral}, unitarity \cite{Marino2014}, and many extensions has been studied in the literature 
\cite{Alves,Alves2,Ozela,Ozela2,VanSergio,MarioPRD2025}. The Proca pseudo-ED also is one of the extensions studied in the literature, 
but it breaks the gauge invariance due to the massive term for the gauge field. The addition of Chern-Simon term yields a more complete description 
to include topological effects in $1+2$ dimensions that preserves the gauge invariance of the theory, with applications in superconductors \cite{Xing}. 
An alternative theory that introduces massive degree of freedom for gauge fields in which the gauge invariance is preserved is known 
as the Lee-Wick electrodynamics (LWED) \cite{lw69,lw70,podolsky42,podolsky44}. The connection of the LWED with topological effects is
a research topic of interest in the literature \cite{accioly03,accioly04,accioly05}, and also the coupling of the LW gauge field 
with fermions \cite{accioly11,turcati14,turcati15,Frasca} provides a quantum electrodynamics with finite quantum corrections at one-loop. 
All these references motivate us to investigations in the Lee-Wick pseudo-electrodynamics (LWPED) added to a non-local Chern-Simons term 
in $1+2$ dimensions as a pseudo-electrodynamics more complete, and with two massive degrees of freedom.  
In this paper, the LWPED that was obtained in the reference \cite{MarioPRD2025} is studied in the presence of a non-local Chern-Simons (CS) topological term.
Thereby, we have the called Lee-Wick-Chern-Simons pseudo-electrodynamics (LWCSPED), that is a non-local and topological ED gauge invariant with
a heavy mass associated with the LW gauge field, and a light mass of the CS topological term. We obtain the gauge propagator, and the corresponding poles that show the massive degree of freedoms of the model. With these results, we calculate the static potential for two like-point charges at rest under the influence of the LW mass, and of the CS parameter. We show that this theory respects the causality principle through the retarded Green function. Posteriorly, we include fermions in the context of Lee-Wick theories. Thereby, the sector of fermions is composite by a light fermion (that is the electron in the material) and the LW fermion (heavy fermion) with the Fermi velocity attached to spatial component of the Dirac matrices in $1+2$ dimensions. We couple the sector of fermions to the LWCSPED gauge field, that we call LWCS pseudo-quantum electrodynamics (LWCSPQED). This model provides a most complete description of the interactions of electrons with a massive non-local gauge field in the presence of topological effects ruled by the CS term. The perturbation theory for the LWCSPQED is showed through the effective action, where the Ward identity must be satisfied as consequence of the gauge invariance. Using this approach, we obtain the radiative corrections to the electron propagator, to the gauge propagator (vacuum polarization), and to the $3$-vertex involving the electron-positron pair and the LWCSPED gauge field. All these quantities are calculated at one loop approximation. The results show the Lee-Wick mass jointly with the $1+2$ dimensions have an important role in the radiative corrections of the model. From the result for the vacuum polarization, the Uehling potential energy 
of two static charges corrected by quantum effects is numerically calculated. The correction at one loop to the $3$-vertex is so calculated via Ward identity since that we know the electron self-energy result. The electron's $g-2$ gyromagnetic factor is obtained in terms of the LW mass and of the CS parameter. We also discuss the unitarity of the LWCS pseudo-ED in the presence of the CS parameter using the Optical theorem.            
The paper is organized as follows : In the section \ref{sec2}, we add a non-local Chern-Simons term to the Lee-Wick pseudo-ED sector investigating 
properties of the model as the potential energy, and the causality principle. 
In the section \ref{sec3}, we couple the sector of fermions with the Lee-Wick-Chern-Simons gauge field to obtain the LWCSPQED. 
We show the effective action approach and the Ward identity in the section \ref{sec4}. In the section \ref{sec5}, we calculate the radiative 
corrections in the LWCSPQED at one loop approximation. The section \ref{sec6} is reserved for the discussion of the unitarity in the LWCSPED. 
For end, the conclusions are presented in the section \ref{sec7}. 
%
%

%
The natural units system $\hbar=c=1$ is used in this paper. 
The metric is $\eta_{\bar{\mu}\bar{\nu}}=\mbox{diag}(+1,-1,-1)$, in which the bar index $\bar{\mu},\bar{\nu}=\{0,1,2\}$
are adopted for vectors and tensors in $1+2$ dimensions.


\section{The Lee-Wick pseudo-electrodynamics with Chern-Simons term}
\label{sec2}
We start this section with a short review of the dimensional reduction in Lee-Wick (LW) ED. 
The LW lagrangian density in the presence of a covariant gauge fixing term in $1+3$ dimensions is 
\begin{eqnarray} \label{LLW}
\mathcal{L}_{LW} &=& -\frac{1}{4} \, F_{\mu\nu}\left(1+\frac{\Box}{M^2}\right)F^{\mu\nu} 
\nonumber \\
&&
-\frac{1}{2\xi} \, \left[  \left(1+\frac{\Box}{M^2} \right) \partial_{\mu}A^{\mu} \right]^2 
- J_{\mu} \, A^{\mu} \; ,
\end{eqnarray}
in which $F_{\mu\nu}=\partial_{\mu}A_{\nu}-\partial_{\nu}A_{\mu}$ is the EM strength field tensor associated 
to the $A^{\mu}$-potential, $M$ is the LW mass, $\xi$ is the gauge fixing parameter, and $J^{\mu}$ is the external source. 
The limit $M \rightarrow \infty$ recovers the usual Maxwell ED with a covariant gauge fixing term. The dimensional 
reduction of (\ref{LLW}) to $1+2$ dimensions is so set by the correspondent action of (\ref{LLW}) written in terms 
of the source $J^{\mu}$ :
\begin{equation}\label{SJ}
S[J_{\mu}]=-\frac{1}{2} \int d^4x \, d^4x^{\prime} \, J_{\mu}(x) \, \Delta^{\mu\nu}(x-x^{\prime}) \, J_{\nu}(x^{\prime}) \; , 
\end{equation}  
where $\Delta^{\mu\nu}(x-x^{\prime})$ is the Green function of the LW theory
\begin{eqnarray}
&&
\Delta^{\mu\nu}(x-x^{\prime}) = \int \frac{d^4k}{(2\pi)^4} \frac{M^2}{k^2(k^2-M^2)} \, \times
\nonumber \\
&&
\times \, \left[ \, \eta^{\mu\nu}-\frac{ k^{\mu} \, k^{\nu} }{k^2} \frac{k^2-M^2(1-\xi)}{k^2-M^2} \, \right] 
e^{ik\cdot(x-x^{\prime})}
\; . \;\;\;\;
\end{eqnarray}
The sources are constrained to satisfy the condition 
\begin{subequations}
\begin{eqnarray}
J^{\bar{\mu}}(x^{\mu}) &=& j^{\bar{\mu}}(x^{\bar{\mu}})\,\delta(z) \; ,
\label{Jmu}
\\
J^{3}(x^{\mu}) &=& 0 \; ,
\label{J3}
\end{eqnarray}
\end{subequations}
in which $x^{\bar{\mu}}=(x_0,x,y)$ sets the coordinates on the 3D Minkowski space-time.
Using this condition in (\ref{SJ}), the action is reduced to $1+2$ dimensions as
\begin{equation}\label{actionEJ3D}
S[j_{\bar{\mu}}]=-\frac{1}{2} \int d^3\bar{x}\,d^3\bar{x}^{\prime} \, j_{\bar{\mu}}(\bar{x})\,\Delta^{\bar{\mu}\bar{\nu}}(\bar{x}-\bar{x}^{\prime}) \, j_{\bar{\nu}}(\bar{x}^{\prime}) \; , 
\end{equation}
where the Green function now is 
\begin{eqnarray}\label{Deltamunureduced}
&&
\Delta^{\bar{\mu}\bar{\nu}}(\bar{x}-\bar{x}^{\prime})
\left. 
=\eta^{\bar{\mu}\bar{\nu}}\int \frac{d^4k}{(2\pi)^4} 
\frac{M^2}{k^2(k^2-M^2)}
\,  e^{ik\cdot(x-x^{\prime})} \right|_{z=z^{\prime}=0}
\nonumber \\
&&
=\frac{\eta^{\bar{\mu}\bar{\nu}}}{2} \int \frac{d^3\bar{k}}{(2\pi)^3} \, e^{i \bar{k} \cdot ( \bar{x}-\bar{x}^{\prime}  ) } 
\left[ \frac{1}{\sqrt{-\bar{k}^2}}-\frac{1}{\sqrt{-\bar{k}^2+M^2}} \right] \; ,
\hspace{0.5cm}
\end{eqnarray}
where we have calculated the $k_{z}$-integration, and the current conservation impose the condition $k_{\bar{\mu}} \, j^{\bar{\mu}}=0$ in the momentum space. 
Therefore, we need to find the lagrangian density in the coordinate space whose action in $1+2$ dimensions yields (\ref{actionEJ3D}), 
and the Green function (\ref{Deltamunureduced}). The solution is given by pseudo-LW lagrangian
\begin{equation}\label{LLW}
{\cal L}_{PLW}=-\frac{1}{4} \, F_{\bar{\mu}\bar{\nu}}N(\bar{\Box})F^{\bar{\mu}\bar{\nu}}
- j_{\bar{\mu}} A^{\bar{\mu}} \; ,
\end{equation}
where the non-local $N(\bar{\Box})$-operator is defined by
\begin{eqnarray}\label{Nop}
N(\bar{\Box})=\frac{2 \, (\bar{\Box}+M^2)}{(\bar{\Box}+M^2)\sqrt{\bar{\Box}}-\bar{\Box}\,\sqrt{\bar{\Box}+M^2}} \; .
\end{eqnarray}
The pseudo-LW ED is $U(1)$ gauge invariance under the transformation $A_{\bar{\mu}} \mapsto A^{\prime}_{\bar{\mu}}
=A_{\bar{\mu}}-\partial_{\bar{\mu}}\Lambda$, in which $\Lambda$ is a real function of the coordinates $x^{\bar{\mu}}=(t,x,y)$,
since that the current satisfies the continuity equation $\partial_{\bar{\mu}}j^{\bar{\mu}}=0$. Thereby, the gauge invariance 
allows us the addition of a topological and non-local Chern-Simons (CS) term given by
\begin{eqnarray}\label{LCS}
{\cal L}_{CS}=\frac{\theta}{2} \, \epsilon^{\bar\mu\bar\nu\bar\rho} A_{\bar{\mu}} N(\bar{\Box}) \, \partial_{\bar{\nu} } A_{\bar\rho} \; .
\end{eqnarray}
Therefore, the LWCS pseudo-ED is defined as the sum of (\ref{LLW}) with (\ref{LCS}) :
%
%
\begin{equation}\label{LLWCS}
{\cal L}_{PLWCS}=-\frac{1}{4} \, F_{\bar{\mu}\bar{\nu}}N(\bar{\Box})F^{\bar{\mu}\bar{\nu}}
+\frac{\theta}{2} \epsilon^{\bar\mu\bar\nu\bar\rho} A_{\bar{\mu}} N(\bar{\Box}) \partial_{\bar{\nu} } A_{\bar\rho}
- j_{\bar{\mu}} A^{\bar{\mu}} \, .
\end{equation}
%
%
Investigations as causality, unitarity, and the quantum corrections at one loop
when the LWCS gauge field is coupled to the fermions are the main goals of this paper.    
The tensors, vectors and parameters from (\ref{LLWCS}) are defined as follows : 
$F_{\bar{\mu}\bar{\nu}}=\partial_{\bar{\mu}}A_{\bar{\nu}}-\partial_{\bar{\nu}}A_{\bar{\mu}}$ is the EM strength field tensor, 
$A^{\bar{\mu}}$ is the gauge field, $\theta$ is the Chern-Simons parameter with mass dimension, and 
$j^{\bar{\mu}}$ is a classical source. As mentioned previously, the bar index runs as $\bar{\mu}=(0,1,2)$, the derivative operator is $\partial_{\bar{\mu}}=(\partial_{t},\partial_{x},\partial_{y})$, and the bar D'Alembertian operator means $\bar{\Box}=\partial_{\bar{\mu}}\partial^{\bar{\mu}}=\partial_{t}^2-\partial_{x}^2-\partial_{y}^2$ in $1+2$ dimensions. 
The limit $M \rightarrow \infty$ reduces the $N$-operator to $N(\bar{\Box})=2/\sqrt{\bar{\Box}}$, and (\ref{LLWCS}) becomes 
the lagrangian of the pseudo-electrodynamics in the presence of a non-local Chern-Simons term 
discussed in the ref. \cite{Alves2}.    
The action principle applied to the lagrangian yields the field equation 
\begin{eqnarray}\label{EqA}
N(\bar{\Box})\,\partial_{\bar{\mu}}F^{\bar{\mu}\bar{\nu}}+\theta\,N(\bar{\Box})\,\tilde{F}^{\bar{\nu}} = j^{\bar{\nu}} \; ,
\end{eqnarray}
in which $\tilde{F}^{\bar{\mu}}=\frac{1}{2}\,\epsilon^{\bar{\mu}\bar{\alpha}\bar{\beta}}F_{\bar{\alpha}\bar{\beta}}$ 
is the dual tensor of $F^{\bar{\mu}\bar{\nu}}$, that satisfies the Bianchi identity $\partial_{\bar{\mu}}\tilde{F}^{\bar{\mu}}=0$.
Thus, we have the field equations for the PLWCS ED. 
%
%
%
%
To obtain the gauge propagator of the PLWCS ED, we add the gauge fixing term to the lagrangian (\ref{LLWCS})
\begin{eqnarray} \label{PLLWresultOpN}
\mathcal{L}_{gf} 
=\frac{1}{2\xi} \, A_{\bar{\mu}}  \left(1+\frac{\bar{\Box}}{M^2} \right) N(\bar{\Box}) \, \partial^{\bar{\mu}}\partial^{\bar{\nu}}A_{\bar{\nu}} \; ,
\end{eqnarray}
where $\xi$ is a real parameter. The modified lagrangian by the gauge fixing can be written as
\begin{eqnarray}
{\cal L}_{PLWCS}=\frac{1}{2} \, A^{\bar{\mu}} \, {\cal O}_{\bar{\mu}\bar{\nu}} \, A^{\bar{\nu}} - j_{\bar{\mu}} \, A^{\bar{\mu}} \; ,
\end{eqnarray}
where the operator ${\cal O}_{\bar{\mu}\bar{\nu}}$ is
\begin{equation}
{\cal O}_{\bar{\mu}\bar{\nu}}=N(\bar{\Box})\bar{\Box}\,\theta_{\bar{\mu}\bar{\nu}}+\frac{1}{\xi}\left(1+\frac{\bar{\Box}}{M^2} \right)N(\bar{\Box})\bar{\Box}\,\omega_{\bar{\mu}\bar{\nu}}
+\theta N(\bar{\Box}) S_{\bar{\mu}\bar{\nu}} \; ,
\end{equation}
and the projectors are given by
\begin{equation}
\theta_{\bar{\mu}\bar{\nu}}=\eta_{\bar{\mu}\bar{\nu}}-\frac{\partial_{\bar{\mu}} \, \partial_{\bar{\nu}}}{\bar{\Box}}
\; , \;
\omega_{\bar{\mu}\bar{\nu}}=\frac{\partial_{\bar{\mu}} \, \partial_{\bar{\nu}}}{\bar{\Box}}
\; , \;
S_{\bar{\mu}\bar{\nu}}=-\epsilon_{\bar{\mu}\bar{\nu}\bar{\rho}}\,\partial^{\bar{\rho}} \; ,
\end{equation}
that satisfy the relations
\begin{eqnarray}
\theta_{\bar{\mu}\bar{\nu}}\,\theta^{\bar{\nu}\bar{\alpha}} &=& \theta_{\bar{\mu}}^{\;\,\,\bar{\alpha}} 
\; , \;
\omega_{\bar{\mu}\bar{\nu}}\,\omega^{\bar{\nu}\bar{\alpha}}=\omega_{\bar{\mu}}^{\;\,\,\bar{\alpha}}
\; , \;
S_{\bar{\mu}\bar{\nu}}\,S^{\bar{\nu}\bar{\alpha}}=-\bar{\Box} \, \theta_{\bar{\mu}}^{\;\,\,\bar{\alpha}} \; ,
\nonumber \\
\theta_{\bar{\mu}\bar{\nu}}\,S^{\bar{\nu}\bar{\alpha}} &=& S_{\bar{\mu}}^{\;\,\,\bar{\alpha}} 
\; , \;
\theta_{\bar{\mu}\bar{\nu}}\,\omega^{\bar{\nu}\bar{\alpha}}=\omega_{\bar{\mu}\bar{\nu}}\,S^{\bar{\nu}\bar{\alpha}}=0
\; .
\end{eqnarray}
Using these relations, the inverse of the ${\cal O}_{\bar{\mu}\bar{\nu}}$ - operator is 
\begin{eqnarray}
({\cal O}_{\bar{\mu}\bar{\nu}})^{-1} &=& \frac{\theta_{\bar{\mu}\bar{\nu}} }{ N(\bar{\Box})(\bar{\Box} +\theta^2)}
+\frac{\xi \, M^2 \, \omega_{\bar{\mu}\bar{\nu}} }{\bar{\Box}\,N(\bar{\Box})(\bar{\Box}+M^2)}
\nonumber \\
&&
-\frac{ \theta \, S_{\bar{\mu}\bar{\nu}}}{ \bar{\Box} \, N(\bar{\Box}) ( \bar{\Box} +\theta^2 )} \; . 
\end{eqnarray}
The functional integration over the $A^{\bar{\mu}}$-field reproduces the effective lagrangian in terms of the source $j^{\bar{\mu}}$
\begin{eqnarray}\label{LLWCSeff}
{\cal L}_{PLWCS}^{eff}[j_{\bar{\mu}}] &=& -\frac{1}{2} \, j^{\bar{\mu}} \, ({\cal O}_{\bar{\mu}\bar{\nu}})^{-1} \, j^{\bar{\nu}}
\nonumber \\
&=& -\frac{1}{2}\,j_{\bar{\mu}}\,\frac{ 1 }{ N(\bar{\Box})(\bar{\Box}+\theta^2)}\,j^{\bar{\mu}}
\nonumber \\
&&
-\frac{\theta}{2} \, \frac{ j^{\bar{\mu}} \, \epsilon_{\bar{\mu}\bar{\nu}\bar{\rho}}\,\partial^{\bar{\rho}}j^{\bar{\nu}} }{ \bar{\Box} \, N(\bar{\Box}) (\bar{\Box} + \theta^2)} \; ,
\end{eqnarray}
in which we have used the continuity equation for current. The second line in (\ref{LLWCSeff}) is interpreted as the electromagnetic 
interaction in the LWCSPED theory. The third line describes the statistical interaction modified by the Lee-Wick mass \cite{Marino93}.  

The gauge propagator is obtained in the momentum space by the correspondence $\partial_{\bar{\mu}} \mapsto -ik_{\bar{\mu}}$, 
such that $\bar{\Box} \mapsto -k_{\bar{\mu}}\,k^{\bar{\mu}}\equiv -\bar{k}^2$, in which the inverse operator 
$({\cal O}_{\bar{\mu}\bar{\nu}})^{-1}$ is written as
\begin{eqnarray}\label{prop}
\Delta_{\bar{\mu}\bar{\nu}}(\bar{k}) &=& \frac{i \, \theta_{\bar{\mu}\bar{\nu}} }{ N(-\bar{k}^2)(\bar{k}^2-\theta^2)}
-\frac{i\, \xi \, M^2 \, \omega_{\bar{\mu}\bar{\nu}} }{\bar{k}^2 \, N(-\bar{k}^2)(\bar{k}^2-M^2)}
\nonumber \\
&&
-\frac{ \theta \, \epsilon_{\bar{\mu}\bar{\nu}\bar{\rho}}\,k^{\bar{\rho}} }{ \bar{k}^2 \, N(-\bar{k}^2) (\bar{k}^2-\theta^2)} \; , 
\end{eqnarray}
and $N(-\bar{k}^2)$ is the operator (\ref{Nop}) in the momentum space, with the projectors $\theta_{\bar{\mu}\bar{\nu}}=\eta_{\bar{\mu}\bar{\nu}}-k_{\bar{\mu}}\,k_{\bar{\nu}}/\bar{k}^2$ 
and $\omega_{\bar{\mu}\bar{\nu}}=k_{\bar{\mu}}\,k_{\bar{\nu}}/\bar{k}^2$. The limit $M \rightarrow \infty$ 
in the result (\ref{prop}) recovers the gauge propagator of the PED added to the non-local CS term. Using the current 
conservation in the momentum space $k_{\bar{\mu}}\,j^{\bar{\mu}}=0$, the contraction of (\ref{prop}) with 
$j^{\bar{\mu}}$ yields
\begin{eqnarray}\label{propJ}
j^{\bar{\mu}}(\bar{k})\,\Delta_{\bar{\mu}\bar{\nu}}(\bar{k}) \, j^{\bar{\nu}}(\bar{k}) &=& \frac{i \, j_{\bar{\mu}}\,j^{\bar{\mu}} }{ N(-\bar{k}^2)(\bar{k}^2-\theta^2)} \; , 
\end{eqnarray}
and using the definition of $N(-\bar{k}^2)$, we can write the previous expression as 
\begin{equation}\label{propJpoles}
j^{\bar{\mu}}(\bar{k})\,\Delta_{\bar{\mu}\bar{\nu}}(\bar{k}) \, j^{\bar{\nu}}(\bar{k}) = 
\frac{j_{\bar{\mu}}\,j^{\bar{\mu}} \, \bar{k}^2}{\bar{k}^2-\theta^2} \left[\frac{-i}{\sqrt{-\bar{k}^2}}+\frac{i}{\sqrt{-\bar{k}^2+M^2}} \right] 
\; .
\end{equation}
%
%
%
This result shows explicitly the poles at $\bar{k}^2 = 0$, $\bar{k}^2 = M^2$, as expected by the LW theory, and also at 
$\bar{k}^2=\theta^2$ that is one consequence of the CS topological term. We use the hierarchy of $M \gg \theta$, in which
theory presents two massive degree of freedoms: a heavy mass $M$, a light mass $\theta$; and a massless degree freedom that 
is consequence of the gauge invariance. 
%
%
%
The propagator has a positive residue if the $3$-current is space-like $j_{\bar{\mu}}\,j^{\bar{\mu}} < 0$. 
%
%
%
%
The limit without the Chern-Simons term $(\theta \rightarrow 0)$ leads 
to the usual poles from the Lee-Wick pseudo-ED. In the ultraviolet regime
in which $\bar{k}^2 \gg M^2$ and $\bar{k}^2 \gg \theta^2$, 
the gauge propagator goes with $\Delta_{\bar{\mu}\bar{\nu}} \sim (\bar{k})^{-1}$. This behavior is the same from PQED, 
but the Lee-Wick mass may works like a natural regulator parameter for divergences in $1+2$ dimensions, as we will see in the section \ref{sec5}. 
%

%
%

%
%

\begin{figure}
\includegraphics[width=\linewidth]{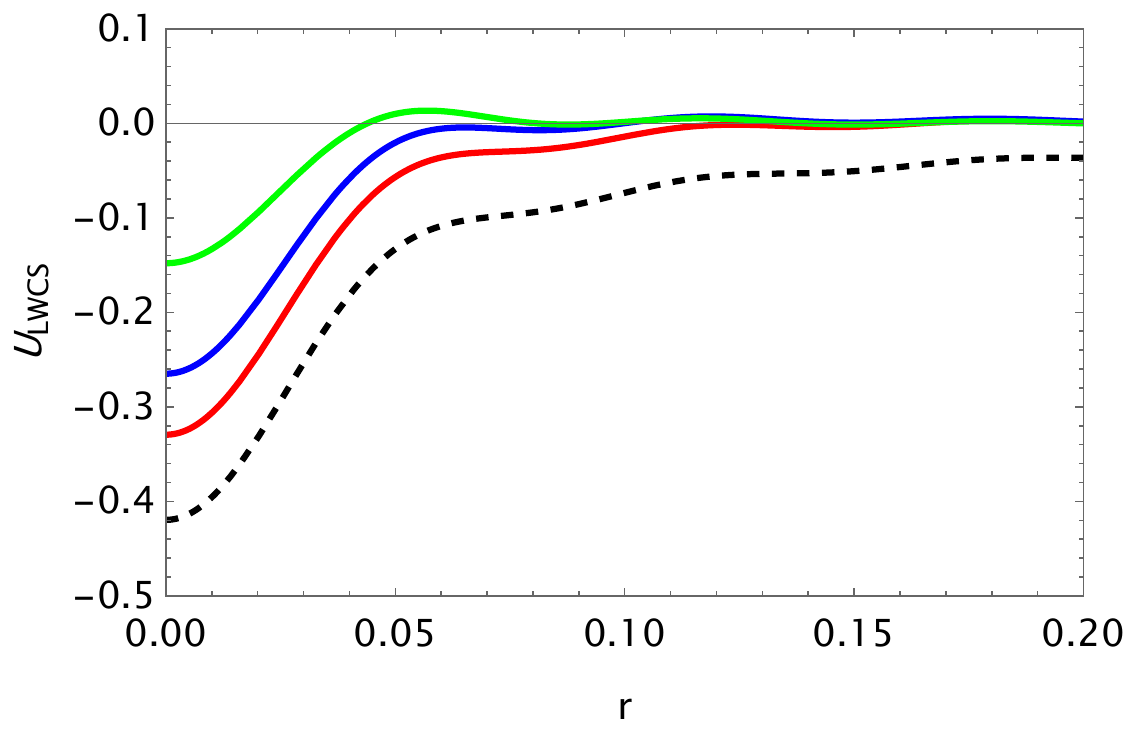}
\caption{The static energy as function of the radial distance. We choose the values $e=0.3$, $M=100$, 
and the colored lines means the $\theta$-values : $\theta=0.0$ (dashed line), $\theta=10$ (red line), $\theta=20$ (blue line) 
and $\theta=50$ (green line).} \label{Fig1}
\end{figure}

%
Backing to the effective lagrangian (\ref{LLWCSeff}), the interaction term for a static like point charge in that 
the current density is $j^{\bar{\mu}}(\bar{{\bf r}})=[-e \, \delta^{2}(\bar{{\bf r}}-{\bf r}),{\bf 0}]$ yields the static 
energy potential between two charges $(-e)$ separated by a distance ${\bf r}$
\begin{equation}\label{Ur}
U({\bf r})=-e^2 \int \frac{d^{2}\bar{{\bf k}}}{(2\pi)^2} \, \frac{ e^{i\,\bar{{\bf k}}\cdot{\bf r}} }{ N(\bar{{\bf k}}^2) (\bar{{\bf k}}^2+\theta^2)} \; ,
\end{equation}
that is reduced to integral
\begin{equation}\label{Urx}
U(r)=-e^2 \int_{0}^{\infty} \frac{dk}{2\pi} \, \frac{k \, J_{0}(k\,r) }{N(k^2)(k^2+\theta^2)} \; ,
\end{equation}
where $r$ is the radial coordinate on the spatial plane, and $J_{0}(k\,r)$ is a Bessel function of the first kind. 
The analytical solution of (\ref{Urx}) for $M \gg \theta$ is given by
\begin{equation}\label{Urtreelevelresult}
U(r)=-\frac{e^2}{4\pi}\left[ \, \frac{1-e^{-Mr}}{r}-\frac{\pi\theta}{2} I_{0}(\theta r)+\frac{\pi\theta}{2} L_{0}(\theta r) \, \right] \; ,
\end{equation}
where $I_{0}$ is a modified Bessel function of first kind, and $L_{0}$ is a Struve function. In the limit $\theta \rightarrow 0$, the usual Lee-Wick potential 
is recovered. The limit $r \rightarrow 0$ yields the finite potential at the origin
\begin{eqnarray}
U(r \rightarrow 0)=-\frac{e^2}{4\pi} \left(M-\frac{\pi\theta}{2}\right)\simeq-\frac{e^2M}{4\pi} \; .
\end{eqnarray}
The numerical solution for the static energy (\ref{Urx}) is plotted as function of the radial distance $(r)$ in the figure (\ref{Fig1}). 
In this plot, we choose $e=0.3$, the Lee-Wick mass as $M=100$, for the values of $\theta=0.0$ 
(this is the case without Chern-Simons term illustrated by the black dashed line), $\theta=10$ (red line), $\theta=20$ (blue line) 
and $\theta=50$ (green line). 
In all the cases the energy is finite at the origin, that is well known in Lee-Wick ED. 
The effect of the $\theta$-parameter is the decreasing of interaction energy at the origin.


%
Two properties are fundamentals for a consistent field theory : causality and unitarity.
The causality of the LW pseudo-ED is confirmed in the ref. \cite{TesePedroOrtega}. In the presence 
of the non-local CS term in (\ref{LLWCS}), this analysis starts with the Green function associated with the propagator 
(\ref{prop}). Choosing the gauge $\xi=0$, the Green function in the coordinate space is given by
\begin{eqnarray}\label{Deltamunu}
\Delta_{\bar{\mu}\bar{\nu}}(\bar{x})= -\frac{1}{2} \, \bar{\Box} \, \hat{P}_{\bar{\mu}\bar{\nu}} \, \Delta(\bar{x}) \; ,
\end{eqnarray}
where $\hat{P}_{\bar{\mu}\bar{\nu}}=\eta_{\bar{\mu}\bar{\nu}}-\partial_{\bar{\nu}}\partial_{\bar{\nu}}/\bar{\Box}
-i\,\theta\,\epsilon_{\bar{\mu}\bar{\nu}\bar{\rho}}\,\partial^{\bar{\rho}}/\bar{\Box}$, and $\Delta(\bar{x})$ is defined by
\begin{equation}\label{intFG}
\Delta(\bar{x})=\int \frac{d^3\bar{k}}{(2\pi)^3} \, \frac{e^{-i \bar{k} \cdot \bar{x}}}{\bar{k}^2-\theta^2} \, \left[ \, \frac{1}{\sqrt{-\bar{k}^2}}-\frac{1}{ \sqrt{-\bar{k}^2+M^2} } \, \right] \; ,
\end{equation}
with $\bar{x}=\bar{x}-\bar{x}^{\prime}$. The advanced $(+)$ and retarded $(-)$ Green functions are obtained by the prescription of $k^{0} \rightarrow k^{0} \mp i\,\epsilon \, (\epsilon > 0)$ in the integral (\ref{intFG}), whose result is given by
\begin{eqnarray}\label{Deltaresult}
\Delta^{\pm}(\bar{x})&=&\frac{i}{4\pi^2} \Theta(\pm x_{0}) \Theta(||\bar{x}||^{2})\left[ G_{13}^{21}\left(\frac{-||\bar{x}||^2\theta^2}{4} \left|^{0}_{0,0,-1/2}  \right. \right) 
\right.
\nonumber \\
&&
\left.
-\pi\,K_{0}(i\,M\sqrt{||\bar{x}||^2})\,H_{-1}(M\sqrt{||\bar{x}||^2})
\right.
\nonumber \\
&&
\left.
-\pi\,K_{1}(i\,M\sqrt{||\bar{x}||^2})\,H_{0}(M\sqrt{||\bar{x}||^2})
\phantom{\frac{1}{2}} \!\!\!\!\right] ,
\end{eqnarray}
where we have used $M \gg \theta$, with $||\bar{x}||^{2}=|x_0|^2-|{\bf x}|^2$, $G_{13}^{21}$ is a Meijer-G function, 
$K_{0}$ and $K_{1}$ are modified Bessel functions of the second kind, $H_{-1}$ and $H_{0}$ are Struve functions. 
Notice that the Heaviside function $\Theta(||\bar{x}||^{2})$ is null outside the light cone, that is $|x_0|^2-|{\bf x}|^2 < 0$. 
Thus, the causality principle is not violated in the pseudo-LW ED with the presence of the non-local Chern-Simons term.
The limit $\theta \rightarrow 0$ in (\ref{Deltamunu})-(\ref{Deltaresult}) recovers the result of the pure pseudo-LW ED. 
\section{The Lee-Wick-Chern-Simons pseudo-quantum electrodynamics}
\label{sec3}

We start this section with the Dirac lagrangian in the Lee-Wick approach
\begin{eqnarray}\label{LDLW}
\mathcal{L}_{D}=\bar{\hat{\psi}}\left(i\,\Gamma^{\bar{\mu}}\partial_{\bar{\mu}}-m\,\mathds{1}\right)\hat{\psi}+\frac{i}{M_{f}^2} \, \bar{\hat{\psi}} \, (\Gamma^{\bar{\mu}}\partial_{\bar{\mu}})^{3} \, \hat{\psi} \; ,
\end{eqnarray} 
where $\hat{\psi}$ is a Dirac spinor in $1+2$ dimensions, the Dirac matrices have the components 
$\Gamma^{\bar{\mu}}=(\gamma^{0},\beta\,\gamma^{i})\,(i=1,2)$, in which $\beta \equiv v_F$ is the Fermi velocity, the $\gamma^{\bar{\mu}}$-matrices 
satisfy the relation $\gamma^{\bar{\mu}}\gamma^{\bar{\nu}}=\eta^{\bar{\mu}\bar{\nu}}+i \, \sigma^{\bar{\mu}\bar{\nu}}$, 
$\bar{\hat{\psi}}=\hat{\psi}^{\dagger}\gamma^{0}$ is the adjunct field. The $\gamma^{\bar{\mu}}$-matrices in the Dirac basis are set by
\begin{eqnarray}
\gamma^{0}&=&\sigma_{3}=
\left[
\begin{array}{cc}
1 & 0 \\
0 & -1 \\
\end{array}
\right]
\; , \;
\gamma^{1}=i\sigma_2=\left[
\begin{array}{cc}
0 & 1 \\
-1 & 0 \\
\end{array}
\right]
\; , \;
\nonumber \\
\gamma^{2}&=&-i\sigma_{1}=\left[
\begin{array}{cc}
0 & -i \\
-i & 0 \\
\end{array}
\right] \; ,
\end{eqnarray}
in which $\sigma_{i}=(\sigma_{1},\sigma_{2},\sigma_{3})$ are the Pauli matrices. The new Lee-wick term introduces the fermion mass 
$M_{f}$ that corresponds to the heavy partner fermion of the electron of mass $m=0.5$ MeV. The limit $M_{f} \rightarrow \infty$ reduces 
(\ref{LDLW}) to the usual lagrangian applied for Dirac materials. The action principle applied in (\ref{LDLW}) yields the Dirac-Lee-Wick 
equation
\begin{eqnarray}\label{EqDLW}
\left(i\,\Gamma^{\bar{\mu}}\partial_{\bar{\mu}}-m\,\mathds{1}\right)\hat{\psi}+\frac{i}{M_{f}^2} 
\, (\Gamma^{\bar{\mu}}\partial_{\bar{\mu}})^{3} \, \hat{\psi}=0 \; .
\end{eqnarray} 
Using the plane wave solution $\hat{\psi}(\bar{x})=u(\bar{p}) \, e^{-i\bar{p}\cdot\bar{x}}$ in (\ref{EqDLW}), 
we obtain
\begin{eqnarray}\label{EqDLWup}
\left[ \, \Gamma^{\bar{\mu}}p_{\bar{\mu}} \left(1-\frac{\bar{p}^2}{M_{f}^2} \right)-m\,\mathds{1} \, \right] u(\bar{p})
=0 \; ,
\end{eqnarray} 
where $\bar{p}^2=p_{\bar{\mu}}p^{\bar{\mu}}=p_{0}^2-\beta^2\,{\bf p}^2$ is the momentum squared in the fermion sector due to the 
$\Gamma^{\bar{\mu}}$-algebra. The adjoint equation of (\ref{EqDLWup}) is
\begin{eqnarray}\label{EqDLWupbar}
\bar{u}(\bar{p})\left[\, \Gamma^{\bar{\mu}}p_{\bar{\mu}} \left( 1- \frac{\bar{p}^2}{M_{f}^2}\right)+m\,\mathds{1} \, \right]
=0 \; ,
\end{eqnarray}
in which $\bar{u}=u^{\dagger}\,\gamma^{0}$. Multiplying the eq. (\ref{EqDLWup}) by $\bar{u}(\bar{p}^{\prime})\,\Gamma^{\bar{\nu}}$ on the left-side, 
the eq. (\ref{EqDLWupbar}) by $\Gamma^{\bar{\nu}}u(\bar{p})$, and summing the two resultants equations, we obtain the components of the 
current in the momentum space 
\begin{subequations}
\begin{eqnarray}
&&
\bar{u}(\bar{p}) \, \gamma^{0} \, u(\bar{p}^{\prime}) = \bar{u}(\bar{p}) \left[ \, \frac{p_{0}+p_{0}^{\prime}}{2m}+i\,\sigma^{0i}\,\frac{\beta \, q_{i}}{2m} 
\right.
\nonumber \\
&&
\left.
-\frac{\bar{p}^{\prime\,2}}{2mM_{f}^2}(p_{0}^{\prime}+i\beta\,\sigma^{0i}p_{i}^{\prime})-\frac{\bar{p}^{2}}{2mM_{f}^2}(p_{0}+i\beta\,\sigma^{0i}p_{i})
\, \right] u(\bar{p}^{\prime}) \; ,
\label{currentu0}
\nonumber \\
\\
&&
\bar{u}(\bar{p}) \, \gamma^{i} \, u(\bar{p}^{\prime}) = \bar{u}(\bar{p}) \left[ \, \beta\,\frac{p^{i}+p^{\prime\,i} }{2m}
+i\frac{\sigma^{i0}q_{0}}{2m}
\right.
\nonumber \\
&&
\left.
+i\,\sigma^{ij}\,\frac{\beta\,q_{j}}{2m}-\frac{\bar{p}^{\prime\,2}}{2mM_{f}^2}(\beta\,p^{\prime\,i}+i\,\sigma^{i0}\,p_{0}^{\prime}
+i\beta\,\sigma^{ij}p_{j}^{\prime})
\right.
\nonumber \\
&&
\left.
-\frac{\bar{p}^{2}}{2mM_{f}^2}(\beta\,p^{i}+i\,\sigma^{i0}\,p_{0}-i\beta\,\sigma^{ij}p_{j}) \, \right] u(\bar{p}^{\prime}) \; ,
\label{currentui}
\end{eqnarray}
\end{subequations}
that is known as the Gordon identity, in which $q^{\bar{\mu}}=p^{\prime\,\bar{\mu}}-p^{\bar{\mu}}$.
For the on-shell condition of $\bar{p}^2=\bar{p}^{\prime2}=m^2$, and $M_{f} \gg m$, the last terms in (\ref{currentu0}) 
and (\ref{currentui}) are neglected. Thereby, the current components for the electron are given by
\begin{subequations}
\begin{eqnarray}
\left. \bar{u}(\bar{p}) \, \gamma^{0} \, u(\bar{p}^{\prime}) \right|_{\bar{p}^2=\bar{p}^{\prime2}=m^2} &\simeq& \bar{u}(\bar{p}) \left[ \, \frac{p_{0}+p_{0}^{\prime}}{2m}+i\,\sigma^{0i}\,\frac{\beta \, q_{i}}{2m} 
\, \right] u(\bar{p}^{\prime}) 
\; , \label{currentu0}
\nonumber \\
\\
\left. \bar{u}(\bar{p}) \, \gamma^{i} \, u(\bar{p}^{\prime}) \right|_{\bar{p}^2=\bar{p}^{\prime2}=m^2} &\simeq& \bar{u}(\bar{p}) \left[ \, \beta\,\frac{p^{i}+p^{\prime\,i} }{2m}
+i\frac{\sigma^{i0}q_{0}}{2m}
\right.
\nonumber \\
&&
\left.
+i\,\sigma^{ij}\,\frac{\beta\,q_{j}}{2m} \, \right] u(\bar{p}^{\prime}) \; .
\label{currentui}
\end{eqnarray}
\end{subequations}
When the on-shell condition is $\bar{p}^2=\bar{p}^{\prime2}=M_{f}^2$, the components are  
\begin{subequations}
\begin{eqnarray}
\left. \bar{u}(\bar{p}) \, \gamma^{0} \, u(\bar{p}^{\prime}) \right|_{\bar{p}^2=\bar{p}^{\prime2}=M_{f}^2} &\simeq& \bar{u}(\bar{p}) \left[ \, -i\,\sigma^{0i}\,\frac{\beta \, p_{i}}{m} 
\, \right] u(\bar{p}^{\prime}) 
\; , 
\label{currentu0M}
\hspace{1cm}
\\
\left. \bar{u}(\bar{p}) \, \gamma^{i} \, u(\bar{p}^{\prime}) \right|_{\bar{p}^2=\bar{p}^{\prime2}=M_{f}^2} &\simeq& \bar{u}(\bar{p}) \left[ 
\, 
-i \, \sigma^{i0} \, \frac{p_{0}}{m}
\, \right] u(\bar{p}^{\prime}) \; .
\label{currentuiM}
\end{eqnarray}
\end{subequations}

%
The fermion propagator in the momentum space is
\begin{eqnarray}\label{propfermion}
&&
S_{F}(p^{\bar{\mu}}) = \frac{i}{\Gamma^{\bar{\mu}}p_{\bar{\mu}}\left(1- \bar{p}^2/M_{f}^2 \right)-m}
\nonumber \\
&&
\simeq  \frac{i}{\gamma^{0}p_{0}+\beta\gamma^{i}p_{i}-m}
-\frac{i}{\gamma^{0}p_{0}+\beta\gamma^{i}p_{i}-M_{f}}  \; ,
\end{eqnarray}
if $M_{f} \gg m$ , whose poles are at $\bar{p}^2 \simeq m^2$ and 
$\bar{p}^2 \simeq M_{f}^2$. In the ultraviolet regime, when $\bar{p}^2 \gg ( \, M_{f}^{2} \,  , \, m^2 \,)$, 
the propagator behaves as $S_{F} \sim (\bar{p})^{-3}$, that helps in the renormalizability of the model in 
$1+2$ dimensions. This propagator includes a usual fermion of mass $m$ summed to the propagator of a heavy fermion 
of mass $M_{f}$ with the minus sign. We call attention for this characteristic of the Lee-Wick fermion sector as follows below.   
The coupling of fermions with the Lee-Wick-Chern-Simons gauge field is introduced substituting the derivative operator 
by the covariant derivative operator in the lagrangian (\ref{LDLW}) : $\partial_{\bar{\mu}} \mapsto D_{\bar{\mu}}=\partial_{\bar{\mu}}+i\,e\,A_{\bar{\mu}}$.
The addition of the gauge sector (\ref{LLWCS}) leads us to the LWCSPQED lagrangian 
\begin{eqnarray}\label{LWPQED}
\mathcal{L}_{PQED}=\bar{\hat{\psi}}\left(i\,\slashed{D}-m\,\mathds{1}\right)\hat{\psi}+\frac{i}{M_{f}^2}\,\bar{\hat{\psi}} \, \slashed{D}\slashed{D}\slashed{D} \, \hat{\psi}
\nonumber \\
-\frac{1}{4}\,F_{\bar{\mu}\bar{\nu}}\,N(\bar{\Box})F^{\bar{\mu}\bar{\nu}}+\frac{\theta}{2} \, \epsilon^{\bar\mu\bar\nu\bar\rho} A_{\bar{\mu}} \, N(\bar{\Box}) \, \partial_{\bar{\nu} } A_{\bar\rho} \; ,
\end{eqnarray}
where $\slashed{D}=\Gamma^{\bar{\mu}}D_{\bar{\mu}}$ is the covariant derivative operator contracted 
with the $\Gamma^{\bar{\mu}}$-matrices, and the coupling constant $(e)$ is dimensionless in $1+2$ dimensions, 
that is related to fine structure constant by $e^2=4\pi \, \beta \, \alpha$, in which $\alpha=1/137$ is the fine structure constant.
This lagrangian is $U(1)$-local gauge invariant, but the fermion Lee-Wick term contains the covariant derivative 
operator in third order. To remedy this puzzle, we use the approach for the auxiliary fermion fields. 
After a diagonalization in the fermion sector, 
the lagrangian (\ref{LWPQED}) is written as
\begin{eqnarray}\label{LWPQEDpsichi}
\mathcal{L}_{PQED}=\bar{\psi}\left(i\,\slashed{D}-m\,\mathds{1}\right)\psi
-\bar{\chi}\left(i\,\slashed{D}-M_{f}\,\mathds{1}\right)\chi 
\nonumber \\
-\frac{1}{4}\,F_{\bar{\mu}\bar{\nu}}\,N(\bar{\Box})F^{\bar{\mu}\bar{\nu}}+\frac{\theta}{2} \, \epsilon^{\bar\mu\bar\nu\bar\rho} \, A_{\bar{\mu}} \, N(\bar{\Box}) \, \partial_{\bar{\nu} } A_{\bar\rho} \; ,
\end{eqnarray}
where now, the $\psi$-field sets the light fermion of mass $m$, that is represented by the electron in the material, and the $\chi$-field 
describes the heavy fermion of mass $M_{f}$, whose the dynamics has a minus sign $(-)$ in relation to $\psi$. 
This sector of the $\chi$-fermion explains the second term of the fermion propagator 
(\ref{propfermion}) with the minus sign. These fermions fields can be treated as independents such that 
propagators of $\psi$ and $\chi$ are given by    
\begin{subequations}
\begin{eqnarray}
S_{F}^{\psi}(p^{\bar{\mu}}) &=& \frac{i}{\gamma^{0}p_{0}+\beta\gamma^{i}p_{i}-m}=i \, \frac{\gamma^{0}p_{0}+\beta\gamma^{i}p_{i}+m }{p_{0}^2-\beta^2\,{\bf p}^2-m^2} \; ,
\nonumber \\
\\ 
S_{F}^{\chi}(p^{\bar{\mu}}) &=&
\frac{-i}{\gamma^{0}p_{0}+\beta\gamma^{i}p_{i}-M_{f}}=-i\,\frac{\gamma^{0}p_{0}+\beta\gamma^{i}p_{i}+M_{f}}{p_{0}^2-\beta^2\,{\bf p}^2-M_{f}^2} \; ,
\nonumber \\
\end{eqnarray}
\end{subequations}
respectively. For the formulation of the perturbation theory, it is convenient to use the lagrangian density in the form (\ref{LWPQEDpsichi}).
The couplings of the fermions with the Lee-Wick gauge field are read as
\begin{eqnarray}\label{Lint}
{\cal L}^{int}=- \, e\,\bar{\psi} \, \Gamma^{\bar{\mu}}A_{\bar{\mu}}\, \psi + \, e\,\bar{\chi} \, \Gamma^{\bar{\mu}}A_{\bar{\mu}}\, \chi \; ,
\end{eqnarray}
%
%
%
%
in which the vertex rules for $\psi$ and $\chi$ are 
$V_{\psi}^{\bar{\mu}}=-i\,e\,\Gamma^{\bar{\mu}}=-ie \, (\gamma^{0},\beta\,\gamma^{i})$ and 
$V_{\chi}^{\bar{\mu}}=+i\,e\,\Gamma^{\bar{\mu}}=+ie \, (\gamma^{0},\beta\,\gamma^{i})$, respectively. 
The contraction of the current $e\,\bar{u}(\bar{p})\,\Gamma^{\bar{\mu}}\,u(\bar{p}^{\prime})$ with $A_{\bar{\mu}}$ yields the result
\begin{eqnarray}\label{currentA}
&&
\bar{u}(\bar{p})\,e\,\Gamma^{\bar{\mu}}A_{\bar{\mu}}\,u(\bar{p}^{\prime})=\bar{u}(\bar{p}) \left[ \, e \, \frac{(p_{0}+p_{0}^{\prime})}{2m}\,A_{0}
\right.
\nonumber \\
&&
\left.
+e\,\beta^2 \, \frac{(p_{i}+p_{i}^{\prime})}{2m}\,A^{i}
-\frac{e\,\beta}{2m}\,\sigma^{0i}\,E_{i}+\frac{e\,\beta^2}{2m} \, ({\bm \sigma} \cdot{\bf B}) \, \right] u(\bar{p}^{\prime}) \, , \;\;
\hspace{0.5cm}
\end{eqnarray}
where we have used that $q_{0}\rightarrow i\partial_{t}$ and $q_{i}\rightarrow -i\partial_{i}$. In (\ref{currentA}), 
we extract the magnetic momentum of electron at tree level
\begin{eqnarray}
{\bm \mu}_{e}=g\left( \frac{e\,\beta^2}{2m} \right){\bf S} \; ,
\end{eqnarray}
with $g=2$ and ${\bf S}={\bm \sigma}/2$, that is modified by the Fermi velocity $(\beta)$ to the squared. 
Since that the LW field contains a heavy massive degree of freedom represented by 
the gauge propagator (\ref{propJpoles}), the signal $(+)$ in the second term of (\ref{propJpoles}) 
implies that the LW particle is instable. Thereby, the theory (\ref{LWPQEDpsichi}) predicts two possible 
decays $A \rightarrow \overline{\psi} \, \psi$ and $A \rightarrow \overline{\chi} \, \chi$, if these processes 
are kinetically allowed. Using the known rules of QFT, the decay width of $A \rightarrow \overline{\psi} \, \psi$ 
is read as
\begin{eqnarray}\label{decaywidth}
\Gamma(A\rightarrow \overline{\psi} \, \psi)=\frac{e^2}{36\beta^2}\left(1-\frac{4m^2}{M^2}\right) \; ,
\end{eqnarray}
that must satisfy the kinetic condition of $M > 2m$. Notice that in $1+2$ dimensions, the decay width is dimensionless in natural units. 
Similarly, the decay width of $A \rightarrow \overline{\chi} \, \chi$ is given by (\ref{decaywidth}) substituting the electron mass $(m)$ 
by the $\chi$-fermion mass $(M_{f})$, if the kinetic condition is $M > 2 \, M_{f}$. The introduction of the auxiliary $\chi$-fermion is 
so important for the perturbative approach that we will see in the next section.

\section{Effective action and the Ward identity}
\label{sec4}
Since the coupling constant in the interactions of (\ref{LWPQEDpsichi}) is very small, 
the perturbative approach of QFT can be used through the generating functional associated with 
the fields $\psi$ , $\chi$ and $A^{\bar{\mu}}$
\begin{widetext}
\begin{equation}\label{FunctionalZ}
Z[\bar{\eta},\eta,\bar{\sigma},\sigma,J^{\bar{\mu}}]=\int \mathcal{D}\bar{\psi} \mathcal{D}\psi \, \mathcal{D}\bar{\chi}\mathcal{D}\chi \, \mathcal{D}A^{\bar{\mu}}
\,\mbox{exp} \left[ i\,S_{LWCS}[\bar{\psi},\psi,\bar{\chi},\chi,A^{\bar{\mu}}]  + i \int d^3x \, (\bar{\eta}\psi + \bar{\psi}\eta+\bar{\sigma}\chi + \bar{\chi}\sigma + J_{\bar{\mu}} A^{\bar{\mu}}  )  \right] \; ,
\end{equation}
\end{widetext}
where $S_{LWCS}$ is the action associated with the lagrangian density (\ref{LWPQEDpsichi}) in the presence of the gauge fixing term (\ref{PLLWresultOpN}), $(\bar{\eta},\eta,\bar{\sigma},\sigma)$ are the sources of the fermions, and $J^{\bar{\mu}}$ the sources of the LWCS gauge field. As usual, the generator functional for connected correlation functions is $W[\bar{\eta},\eta,\bar{\sigma},\sigma,J^{\bar{\mu}}]=-i\,\mbox{ln}Z[\bar{\eta},\eta,\bar{\sigma},\sigma,J^{\bar{\mu}}]$. The effective 
action is defined by
\begin{eqnarray}
\Gamma[ \bar{\psi},\psi,\bar{\chi},\chi,A^{\bar{\mu}} ] &=& W[\bar{\eta},\eta,\bar{\sigma},\sigma,J^{\bar{\mu}}]  
\nonumber \\ 
&&
\hspace{-2.0cm}
- \int d^3x \, \left( \, \bar{\eta}\psi + \bar{\psi}\eta + \bar{\sigma}\chi + \bar{\chi}\sigma + J_{\bar{\mu}} A^{\bar{\mu}} \, \right) 
\; ,
\end{eqnarray}
in which we can obtain the relations
\begin{eqnarray}\label{relWGamma}
\frac{\delta W}{\delta J^{\bar{\mu}}} &=& A_{\bar{\mu}}
\;  , \; 
\frac{\delta W}{\delta \eta}=\bar{\psi}
\; , \;
\frac{\delta W}{\delta \bar{\eta}}=\psi 
\; , \;
\frac{\delta W}{\delta \sigma}=\bar{\chi} \, ,
\nonumber \\
\frac{\delta W}{\delta \bar{\sigma}} &=& \chi 
\; , \;
\frac{\delta \Gamma}{\delta A^{\bar{\mu}}}=-J_{\bar{\mu}} 
\; , \;
\frac{\delta \Gamma}{\delta \psi}=-\bar{\eta} \, ,
\nonumber \\
\frac{\delta \Gamma}{\delta \bar{\psi}} &=& -\eta 
\; , \;
\frac{\delta \Gamma}{\delta \chi}=-\bar{\sigma}
\; , \;
\frac{\delta \Gamma}{\delta \bar{\chi}}=-\sigma \, .
\end{eqnarray}

Under the infinitesimal gauge transformation $\psi \mapsto \psi^{\prime}=\psi+ie\,\Lambda$, 
$\chi \mapsto \chi^{\prime}=\chi+ie\,\Lambda$ and $A_{\bar{\mu}} \mapsto A_{\bar{\mu}}^{\prime}=A_{\bar{\mu}}-\partial_{\bar{\mu}}\Lambda$, the gauge fixing and sources terms are not invariants in the generator functional (\ref{FunctionalZ}) for an arbitrary function $\Lambda$. The gauge invariance is recovered if the generating functional satisfies the equation       
\begin{eqnarray}\label{EqZ}
\left[-\frac{i}{\xi}\left(1+\frac{\bar{\Box}}{M^2} \right)N(\bar{\Box})\bar{\Box}\partial^{\bar{\mu}}\frac{\delta}{\delta J^{\bar{\mu}}}+\partial_{\bar{\mu}}J^{\bar{\mu}}
\right.
\nonumber \\
\left.
+e\left( \bar{\eta}\,\frac{\delta}{\delta\bar{\eta}}- \eta\,\frac{\delta}{\delta\eta} \right)
+e\left( \bar{\sigma}\,\frac{\delta}{\delta\bar{\sigma}}- \sigma\,\frac{\delta}{\delta\sigma} \right)
\right]Z=0 \; , \;
\end{eqnarray}
that in terms of $W$-functional is
\begin{eqnarray}\label{EqW}
\frac{1}{\xi}\left(1+\frac{\bar{\Box}}{M^2} \right)N(\bar{\Box})\,\bar{\Box}\,\partial^{\bar{\mu}}\frac{\delta W}{\delta J^{\bar{\mu}}}+\partial_{\bar{\mu}}J^{\bar{\mu}}
\nonumber \\
+ie \bar{\eta}\,\frac{\delta W}{\delta\bar{\eta}}- ie \eta\,\frac{\delta W}{\delta\eta}
+ie \bar{\sigma}\,\frac{\delta W}{\delta\bar{\sigma}}- ie\sigma\,\frac{\delta W}{\delta\sigma} =0 \; . \;\;\;
\end{eqnarray}
Using the relations (\ref{relWGamma}), the functional equation (\ref{EqW}) in terms of the effective action is given by
\begin{eqnarray}\label{EqGamma}
\frac{1}{\xi}\left(1+\frac{\bar{\Box}}{M^2} \right)N(\bar{\Box})\,\bar{\Box}\,\partial_{\bar{\mu}}A^{\bar{\mu}}
-\partial_{\bar{\mu}}\frac{\delta \Gamma}{\delta A_{\bar{\mu}}}
\nonumber \\
+ie\psi\,\frac{\delta \Gamma}{\delta\psi}- ie\bar{\psi}\,\frac{\delta \Gamma}{\delta\bar{\psi}}
+ie\chi\,\frac{\delta \Gamma}{\delta\chi}- ie\bar{\chi}\,\frac{\delta \Gamma}{\delta\bar{\chi}}=0 \; . \;\;\;
\end{eqnarray}
From (\ref{EqGamma}) it is possible to obtain relations between the functions of $2$- and $3$-points for the fermions $\psi$ and $\chi$. 
Taking the functional derivatives of (\ref{EqGamma}) in relation to $\psi(y_{1})$ and $\bar{\psi}(x_{1})$, and after it, we set 
$\bar{\psi}=\psi=\bar{\chi}=\chi=A^{\bar{\mu}}=0$, the equations reads as 
\begin{eqnarray}
-\partial_{x}^{\bar{\mu}}\frac{\delta^3\Gamma[0]}{\delta\bar{\psi}(x_1) \delta\psi(y_1) \delta A^{\bar{\mu}}(x)} &=& ie\delta(x-x_{1})\frac{\delta^2\Gamma[0]}{\delta\bar{\psi}(x_1) \delta\psi(y_1)}
\nonumber \\
&&
\hspace{-1cm}
-ie\delta(x-y_{1})\frac{\delta^2\Gamma[0]}{\delta\bar{\psi}(x_1) \delta\psi(y_1)} \; .
\end{eqnarray}
Similarly, the same step is implemented for the fields $\chi(y_{1})$ and $\bar{\chi}(x_{1})$ in which we obtain the relation 
\begin{eqnarray}
-\partial_{x}^{\bar{\mu}}\frac{\delta^3\Gamma[0]}{\delta\bar{\chi}(x_1) \delta\chi(y_1) \delta A^{\bar{\mu}}(x)} &=& ie\delta(x-x_{1})\frac{\delta^2\Gamma[0]}{\delta\bar{\chi}(x_1) \delta\chi(y_1)}
\nonumber \\
&&
\hspace{-1cm}
-ie\delta(x-y_{1})\frac{\delta^2\Gamma[0]}{\delta\bar{\chi}(x_1) \delta\chi(y_1)} \; .
\end{eqnarray}
The Fourier transform leads to following relations in the momentum space
\begin{subequations}
\begin{eqnarray}
q^{\bar{\mu}}\Gamma_{\bar{\mu}}^{\psi}(p,q,p+q) &=& (S_{full}^{\psi})^{-1}(p+q)-(S_{full}^{\psi})^{-1}(p) 
\, , \hspace{0.8cm}
\\
q^{\bar{\mu}}\Gamma_{\bar{\mu}}^{\chi}(p,q,p+q) &=& (S_{full}^{\chi})^{-1}(p+q)-(S_{full}^{\chi})^{-1}(p) 
\, , \hspace{0.8cm}
\end{eqnarray}
\end{subequations}
where $(S_{full}^{\psi})^{-1}$ and $(S_{full}^{\chi})^{-1}$ are the inverse of full propagators for the fermions 
$\psi$ and $\chi$, respectively, $\Gamma_{\bar{\mu}}^{\psi}$ and $\Gamma_{\bar{\mu}}^{\chi}$ denotes the full 
$3$-point vertex for the couplings (\ref{Lint}). In the limit $q^{\bar{\mu}} \rightarrow 0$, the previous equations yield the 
Ward identities  
\begin{subequations}
\begin{eqnarray}
\Gamma_{\bar{\mu}}^{\psi}(p,0,p) &=& \frac{\partial}{\partial p^{\bar{\mu}}}(S_{full}^{\psi})^{-1} \, ,
\\
\Gamma_{\bar{\mu}}^{\chi}(p,0,p) &=& \frac{\partial}{\partial p^{\bar{\mu}}}(S_{full}^{\chi})^{-1} \, ,
\end{eqnarray}
\end{subequations}
that are valid for all orders in the perturbation theory. The inverse of full propagator in terms of the 
free is $(S_{full}^{\psi(\chi)})^{-1}=(S^{\psi(\chi)})^{-1}-\Sigma^{\psi(\chi)}$, where $\Sigma^{\psi}$ and $\Sigma^{\chi}$ 
are the self-energies for fermions. Writing $\Gamma_{\bar{\mu}}^{\psi(\chi)}(p,q,p+q)=\Gamma_{\bar{\mu}}+\Lambda_{\bar{\mu}}^{\psi(\chi)}(p,q,p+q)$, 
in which $\Gamma_{\bar{\mu}}=(\gamma_{0},\beta\,\gamma_{i})$ are the Dirac matrices components, the ward identities (with $q^{\bar{\mu}}\rightarrow 0$) 
lead to
\begin{subequations}
\begin{eqnarray}
\Lambda_{\bar{\mu}}^{\psi}(p,0,p) &=& -\frac{\partial \Sigma^{\psi}}{\partial p^{\bar{\mu}}} \; ,
\label{Lambdapsi}
\\
\Lambda_{\bar{\mu}}^{\chi}(p,0,p) &=& -\frac{\partial \Sigma^{\chi}}{\partial p^{\bar{\mu}}} \; ,
\label{Lambdachi}
\end{eqnarray}
\end{subequations}
that relates the self-energies with the $3$-vertex corrections for all the orders of the perturbative theory. 
These identities will be useful in the next section.
\section{Radiative corrections at one loop in CSLW PQED}
\label{sec5}
Considering all the quantum corrections, the full electron propagator is written as
\begin{eqnarray}
S_{full}^{\psi}(p^{\bar{\mu}})=\frac{i}{\gamma^{0}p_{0}+\beta\,\gamma^{i}p_{i}-m-\Sigma(\bar{p})} \; ,
\end{eqnarray}
in which the electron's physical mass is redefined by $m^{(p)}=m+\Sigma(\bar{p})$, that is the new propagator pole.
The electron self-energy expression at one-loop approximation from the lagrangian (\ref{LWPQED}) is given by
\begin{equation}\label{ese}
-\, i \, \Sigma_1^{\psi}(\bar{p}) = \int\frac{d^{3}\bar{k}}{(2\pi)^{3}} \, V_{ \psi }^{\bar{\mu}} \, 
\Delta_{\bar{\mu}\bar{\nu}}(\bar{k}) \, S_{F}^{\psi}(\bar{p}-\bar{k})\, V_{\psi}^{\bar{\nu}} 
\, .
\end{equation}
%
%
%
The gauge and fermion propagators in the ultraviolet regime, the power counting in this $\bar{k}$-integral has the degree of divergence 
$D=3-1-1=1$, and shows that the integral (\ref{ese}) is divergent in three dimensions. Therefore, we need to introduce the dimensional 
regularization $(D)$ to make the integral (\ref{ese}) finite. To simplify the calculus in this section, we consider 
the gauge propagator (\ref{prop}) in the Landau gauge $(\xi = 0)$ in (\ref{prop}), we have
\begin{eqnarray}\label{propapprox}
\Delta_{\bar{\mu}\bar{\nu}}(\bar{k}) &=& \frac{- \, i \, \eta_{\bar{\mu}\bar{\nu}} \, \bar{k}^2 }{2 \, (\bar{k}^2-\theta^2)} \left[ \frac{1}{\sqrt{-\bar{k}^2}} - \frac{1}{\sqrt{-\bar{k}^2+M^2}} \right]
\nonumber \\
&&
\hspace{-0.5cm}
+ \frac{\theta \, \epsilon_{\bar{\mu}\bar{\nu}\bar{\rho} } \, k^{\bar{\rho}} }{2\,(\bar{k}^2-\theta^2)} \, \left[ \frac{1}{ \sqrt{-\bar{k}^2} } - \frac{1}{ \sqrt{ -\bar{k}^2+M^2} } \right] \, , \;\;\;\;
\end{eqnarray}
and also we neglect the effects of the $\theta$-parameter in relation to LW and electron masses in the electron self-energy calculus.
Thereby, we consider the hierarchy condition for masses $M \gg m \gg \theta$, in which $\theta$ is very small when compared to $\bar{k}$.   
Under these conditions, the self-energy with the $D$-dimensional regulator can be written as
\begin{eqnarray}
\Sigma_1(\bar{p},D)=\Sigma_1^{LW}(\bar{p},D)+\Sigma_{1}^{CS}(\bar{p},D) \; ,
\end{eqnarray}
where 
\begin{eqnarray}\label{Sigma1LW}
\Sigma_1^{LW}(\bar{p},D) &=& -\frac{ie^2}{2}\int\frac{d^{D}\bar{k}}{(2\pi)^{D}}  \left[ \frac{1}{ \sqrt{ -\bar{k}^2 } } - \frac{1}{ \sqrt{ -\bar{k}^2 +M^2 } }  \right] 
\nonumber \\
&&
\times \, 
\, \frac{\Gamma^{\bar{\mu}} \, [ \, \Gamma^{\bar{\alpha}}(p_{\bar{\alpha}}-k_{\bar{\alpha}})+m \, ] \, \Gamma_{\bar{\mu}}}{ (p_0-k_0)^2-\beta^2 \, ({\bf p}-{\bf k})^2-m^2 }  \; ,
\end{eqnarray}
and 
\begin{eqnarray}\label{Sigma1CS}
\Sigma_1^{CS}(\bar{p},D) &=& -\frac{i\theta}{2} e^2 \int\frac{d^{D}\bar{k}}{(2\pi)^{D}} 
\left[ \, \frac{1}{\sqrt{ -\bar{k}^2 } }
- \frac{1}{ \sqrt{ -\bar{k}^2 + M^2 } } \right] 
\nonumber \\
&&
\times \, 
\frac{\epsilon_{\bar{\mu}\bar{\rho}\bar{\nu}}\,k^{\bar{\rho}}\,\Gamma^{\bar{\mu}}  \, [ \, \Gamma^{\bar{\alpha}}(p_{\bar{\alpha}}-k_{\bar{\alpha}})+m \, ] \, \Gamma^{\bar{\nu}}}{ \bar{k}^2 \, [ \, (p_0-k_0)^2-\beta^2 \, ({\bf p}-{\bf k})^2-m^2 \, ]} \, . \;\;\;\;\;\;
\end{eqnarray}
These integrals are calculated by means of the technical of loop integrals from QFT \cite{Peskin}. The general formula for Feynman's parametrization 
\begin{eqnarray}\label{ID}
\frac{1}{A_{1}^{m_1} \cdots A_{n}^{m_n}}= \frac{ \Gamma(m_{1}+\cdots+m_{n} ) }{ \Gamma(m_{1}) \cdots \Gamma(m_{n}) } \int_{0}^{1} dx_{1} \cdots dx_{n} \,
\nonumber \\
 \times \, \delta \left[ \sum_{i=1}^{n}x_i-1 \right]  \frac{ \prod_{i=1}^{n} \, x_{i}^{m_{i}-1} }{ \left[ \, x_{1} \, A_{1} + \cdots + x_{n} \, A_{n}  \, \right]^{m_1 + \cdots + m_n} } \, , 
 \hspace{0.5cm}
\end{eqnarray}
can be used in (\ref{Sigma1LW}) and (\ref{Sigma1CS}), when $m_{i} \, (i=1,2,\cdots,n)$ are not integers numbers. After many calculus, the integral (\ref{Sigma1LW}) 
in the limit $D \rightarrow 3$ yields the finite result
\begin{widetext}
\begin{eqnarray}\label{se}
\Sigma_{1}^{LW}(p_{0},p_{i}) &=& \frac{e^2}{32\pi}\int^{1}_{0} \frac{dx}{\sqrt{1-x}}\frac{1}{1-(1-\beta^2)x}  \left\{ m(1+2\beta^2)-(2\beta^2-1)(1-x)\gamma^{0}p_{0}
-\frac{(1-x)\,\beta\,\gamma^{i}p_{i} }{1-(1-\beta^2)x}\right\}
\nonumber\\
&&
\times \, \mbox{ln}\left[\frac{ m^2 \, x-p_{0}^2 \, x(1-x)+\beta^2{\bf p}^2x(1-x)(1-(1-\beta^2)x)^{-1} }{ M^2(1-x) + m^2 \, x-p_{0}^2 \, x(1-x)+\beta^2{\bf p}^2x(1-x)(1-(1-\beta^2)x)^{-1} }\right] \; , 
\end{eqnarray}
\end{widetext}
in which (\ref{se}) requires the conditions :
\begin{subequations}\label{conditionsInts}
\begin{eqnarray}
2(M^2+m^2)>p_{0}^2-\frac{2\beta^4 \, {\bf p}^2}{1+\beta^2} \; ,
\\
2\,m^2>p_{0}^2-\frac{2\beta^4 \, {\bf p}^2}{1+\beta^2} \; .
\end{eqnarray}
\end{subequations}
The calculus of the integral (\ref{Sigma1CS}) starts with the identity
\begin{eqnarray}
\gamma^{\bar{\mu}}\,\gamma^{\bar{\nu}}\,\gamma^{\bar{\alpha}}=\epsilon^{\bar{\mu}\bar{\nu}\bar{\alpha}}
-\eta^{\bar{\mu}\bar{\nu}}\,\gamma^{\bar{\alpha}}
-\eta^{\bar{\nu}\bar{\alpha}}\,\gamma^{\bar{\mu}}
+\eta^{\bar{\mu}\bar{\alpha}}\,\gamma^{\bar{\nu}} \; ,
\end{eqnarray}
in which it can be written as
\begin{eqnarray}\label{Sigma1CSsimp}
\Sigma_1^{CS}(\bar{p},D) &=& -\frac{i\theta}{2} \, e^2 \! \int\frac{d^{D}\bar{k}}{(2\pi)^{D}} 
\frac{ p_{\bar{\mu}}k^{\bar{\mu}}-k_{\bar{\mu}}k^{\bar{\mu}}+m\,\Gamma^{\bar{\mu}}\,k_{\bar{\mu}} }{ (p_0-k_0)^2-\beta^2 \, ({\bf p}-{\bf k})^2-m^2 } 
\,  
\nonumber \\
&&
\hspace{-0.5cm}
\times \,
\frac{1}{\bar{k}^2} 
\left[ \, \frac{1}{\sqrt{-\bar{k}^2}} - \frac{1}{ \sqrt{ -\bar{k}^2 + M^2 } } \right] \, .
\end{eqnarray}
Using the formula (\ref{ID}), the integral (\ref{Sigma1CSsimp}) in the limit $D \rightarrow 3$ is given by
\begin{widetext}
\begin{eqnarray}
\Sigma_1^{CS}(p_{0},p_{i}) = \frac{\theta\,e^2}{32\pi}\,\int_{0}^{1} \frac{dx}{1-(1-\beta^2)x}\left[1+\frac{2\beta^2}{1-(1-\beta^2)x}\right]
\nonumber \\
\times \left[ \, \sqrt{ M^2(1-x)+m^2\,x-p_{0}^2\,x(1-x)+\beta^2\,{\bf p}^2x\left( \frac{1-x-(1-\beta^2)x}{1-(1-\beta^2)x} \right) }
\right.
\nonumber \\
\left. 
+ \sqrt{ m^2\,x-p_{0}^2\,x(1-x)+\beta^2\,{\bf p}^2x\left( \frac{1-x-(1-\beta^2)x}{1-(1-\beta^2)x} \right) } \, \right]
\nonumber \\
+\frac{\theta\,e^2}{32\pi}\,\int_{0}^{1} \frac{dx}{1-(1-\beta^2)x}\left[ \, p_{0}^2\,x(1-x)-\frac{\beta^2\,{\bf p}^2\,x(1-x)}{(1-(1-\beta^2)x)^2}
+m\gamma^{0}p_{0}\,x+\frac{m\,\beta^2\, \gamma^{i}p_{i}x}{1-(1-\beta^2)x} \, \right]
\nonumber \\
\times \left[ \frac{1}{ \sqrt{ M^2(1-x)+m^2\,x-p_{0}^2\,x(1-x)+\beta^2\,{\bf p}^2x\left( \frac{1-x-(1-\beta^2)x}{1-(1-\beta^2)x} \right) }}
\right.
\nonumber \\
\left.
+\frac{1}{ \sqrt{ m^2\,x-p_{0}^2\,x(1-x)+\beta^2\,{\bf p}^2x\left( \frac{1-x-(1-\beta^2)x}{1-(1-\beta^2)x} \right) }  } \right]
\nonumber \\
-\frac{\theta\,e^2}{16\pi}\,\int_{0}^{1} \frac{dx}{ 1-(1-\beta^2)x }\left[1+\frac{2\beta^2}{1-(1-\beta^2)x}\right] \int_{0}^{1-x} \frac{dy}{\sqrt{y \, (1-x-y)}}
\nonumber \\
\times \sqrt{ M^2(1-x-y)+m^2x-p_{0}^2\,x(1-x)+\beta^2\,{\bf p}^2\,x\left( \frac{1-x-(1-\beta^2)x}{1-(1-\beta^2)x} \right) }
\nonumber \\
-\frac{\theta\,e^2}{16\pi}\,\int_{0}^{1} \frac{dx}{1-(1-\beta^2)x} \int_{0}^{1-x}  \frac{dy}{\sqrt{y\,(1-x-y)}} 
\nonumber \\
\times \left[ \, p_{0}^2\,x(1-x)-\frac{\beta^2\,{\bf p}^2\,x(1-x)}{(1-(1-\beta^2)x)^2}
+m\gamma^{0}p_{0}\,x+\frac{m\,\beta^2\, \gamma^{i}p_{i}x}{1-(1-\beta^2)x} \, \right]
\nonumber \\
\times \frac{1}{ \sqrt{ M^2(1-x-y)+m^2x-p_{0}^2\,x(1-x)+\beta^2\,{\bf p}^2x\left( \frac{1-x-(1-\beta^2)x}{1-(1-\beta^2)x} \right) }}
\; ,
\end{eqnarray}
\end{widetext}
where the same conditions from (\ref{conditionsInts}) must be satisfied.
%


%
When $\beta =1$, the result of the electron self-energy in $3D$ is recovered in the presence of a Chern-Simons term. 
Notice also that the Lee-Wick mass works as regulator parameter, that when $M \rightarrow \infty$, 
the result for the electron self-energy diverges. The electron's self-energy in pure Lee-Wick in pseudo-QED also is recovered in the limit $\theta \rightarrow 0$.  
In particular, if we choose the on-shell external momentum $p^{0}=m$ and $p^{i}=0$ as a condition for the renormalization of the theory, 
the electron self-energy at one loop is reduced to  
\begin{eqnarray}\label{Sigma1Result}
&&
\Sigma_{1}^{\psi}(m) = \frac{e^2\,m}{32\pi}\int^{1}_{0} \frac{dx}{\sqrt{1-x}}\frac{1}{1-(1-\beta^2)x}  
\nonumber \\
&&
\left[(1+2\beta^2)\mathds{1}-(2\beta^2-1)(1-x)\gamma^{0}\right]
\mbox{ln}\left[\frac{ x^2 }{ \mu^2(1-x) + x^2 }\right]
\nonumber \\
&&
+\frac{\theta\,e^2\,m}{32\pi}\,\int_{0}^{1} \frac{dx}{1-(1-\beta^2)x}\left[1+\frac{2\beta^2}{1-(1-\beta^2)x}\right]
\nonumber \\
&&
\times \left[ \, \sqrt{ \mu^2(1-x)+x^2} + x \, \right]
\nonumber \\
&&
+\frac{\theta\,e^2\,m}{32\pi}\,\int_{0}^{1} dx \;  \frac{ (1-x)\mathds{1}
+\gamma^{0} }{1-(1-\beta^2)x}
\left[ 1+\frac{x}{ \sqrt{ \mu^2(1-x)+x^2 }} \right]
\nonumber \\
&&
-\frac{\theta\,e^2\,m}{8\pi}\,\int_{0}^{1} \frac{dx}{ 1-(1-\beta^2)x }\left[1+\frac{2\beta^2}{1-(1-\beta^2)x}\right] 
\nonumber \\
&&
\times 
\sqrt{x^2+(1-x)\mu^2}\,E\left[ \frac{(1-x)\mu^2}{x^2+(1-x)\mu^2} \right]
\nonumber \\
&&
-\frac{\theta\,e^2\,m^2}{8\pi M}\,\int_{0}^{1} \frac{dx}{\sqrt{1-x}} \, \frac{x(1-x)\mathds{1}+\gamma^{0}\,x}{1-(1-\beta^2)x} 
\nonumber \\
&&
\times 
\left\{ \, K\left[ 1+\frac{x^2}{(1-x)\mu^2} \right] + i \, K\left[\frac{-\,x^2}{(1-x)\mu^2} \right] \, \right\} \; ,
\end{eqnarray}
%
%
%
%
where $E$ and $K$ are elliptic functions, and we have defined the dimensionless parameter $\mu:=M/m$. Therefore, the physical electron mass 
is finite, and the fermionic propagator pole is so corrected by the result (\ref{Sigma1Result}).  
The next important contribution at the one-loop is in the gauge field propagator, that is known in QED as the vacuum polarization. 
In the one-loop approximation, the gauge propagator has just the contribution of fermion loop, that in the case of lagrangian 
(\ref{LWPQEDpsichi}), we must add the contributions of $\psi$- and $\chi$-fields. Using the previous rules in the momentum space, the 
vacuum polarization is set by the tensor :   
%
%
%
%
\begin{eqnarray}\label{Polvacuo1loop}
&&
i\,\Pi_{1}^{\bar{\mu}\bar{\nu}}(\bar{k}) = -\int\frac{d^3\bar{p}}{(2\pi)^3} \, \mbox{tr}\left[ \, V_{\psi}^{\bar{\mu}} \, S_{F}^{\psi}(\bar{p})
\, V_{\psi}^{\bar{\nu}} \, S_{F}^{\psi}(\bar{p}-\bar{k}) \right] 
\nonumber \\
&&
-\int\frac{d^3\bar{p}}{(2\pi)^3} \, \mbox{tr}\left[ \, V_{\chi}^{\bar{\mu}} \, S_{F}^{\chi}(\bar{p})
\, V_{\chi}^{\bar{\nu}} \, S_{F}^{\chi}(\bar{p}-\bar{k}) \right]
\; ,
\end{eqnarray}
where the minus sign $(-)$ means the fermion loop in $\psi$- and $\chi$-fields, respectively.

\begin{figure}
\includegraphics[width=\linewidth]{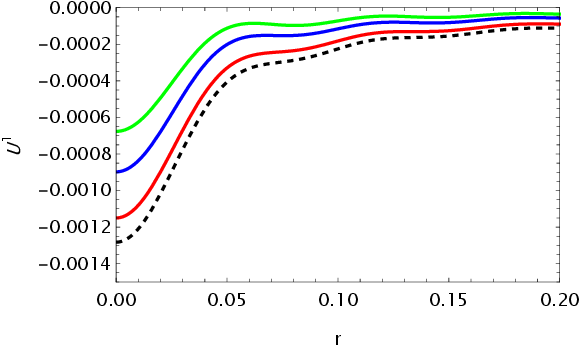}
\caption{The static potential energy corrected by the vacuum polarization at one loop as function of the radial distance. We choose the values of $m=0.5$, $M_{f}=50$, $M=100$, 
and $\beta=0.003$, in which the colored lines means the $\theta$-values : $\theta=0$ (dashed line), $\theta=1.0$ (red line), $\theta=2.0$ (blue line) 
and $\theta=3.0$ (green line).} \label{Fig2}
\end{figure}

%
%
After the trace calculus on the $\Gamma^{\bar{\mu}}$-matrices, we use the technics for loop integrals of QFT \cite{Peskin}. 
The spatial symmetry breaking with the presence of Fermi velocity $(\beta)$ yields the results for the components of 
the vacuum polarization tensor :     
\begin{subequations}
\begin{eqnarray}
\Pi_{1}^{00}(k^{0},{\bf k}) &=& -\frac{e^2}{\pi} \, {\bf k}^2 \, \Pi_1(\bar{k}^2) \; ,
\label{Pi100}
\\
\Pi_{1}^{ij}(k^{0},{\bf k}) &=& \frac{e^2}{\pi} \left[\,\bar{k}^2 \, \eta^{ij}-\beta^2 \, k^{i} \, k^{j} \,\right] \, \Pi_1(\bar{k}^2) \; ,
\\
\Pi_{1}^{0i}(k^{0},{\bf k}) &=& -\frac{e^2}{\pi} \, k^{0} \, k^{i} \, \Pi_1(\bar{k}^2) \; ,
\label{Pi10i}
\end{eqnarray}
\end{subequations}
where the scalar function $\Pi_{1}(\bar{k}^2)$ is given by the integral 
\begin{eqnarray}\label{Pi1int}
\Pi_{1}(\bar{k}^2)=\int_{0}^{1} dx \, \frac{x(1-x)}{\sqrt{m^2-\overline{k}^{\,2} x(1-x)}}
\nonumber \\
+\int_{0}^{1} dx \, \frac{x(1-x)}{\sqrt{M_{f}^2-\overline{k}^{\,2} x(1-x)}} \; ,
\end{eqnarray}
and $\bar{k}^2=k_{0}^2-\beta^2 \, {\bf k}^2$ is the photon external momentum. If we consider the conditions of 
$4M_{f}^2>4m^2>\bar{k}^2$, the result of (\ref{Pi1int}) is  
%
\begin{eqnarray}\label{Pi}
\Pi_1(\bar{k}^2) &=& \frac{1}{4\sqrt{\bar{k}^2}}\left(1+\frac{4m^2}{\bar{k}^2} \right) \coth^{-1}\!\left(\frac{2m}{\sqrt{\bar{k}^2}}\right) 
-\frac{m}{2\bar{k}^2}
\nonumber \\
&&
\hspace{-0.5cm}
+\frac{1}{4\sqrt{\bar{k}^2}}\left(1+\frac{4M_{f}^2}{\bar{k}^2} \right) \coth^{-1}\!\left(\frac{2M_{f}}{\sqrt{\bar{k}^2}}\right) 
-\frac{M_{f}}{2\bar{k}^2}
\; .
\hspace{0.6cm}
\end{eqnarray}
%
%
%
%
Notice that the last line in (\ref{Pi}) is zero in the limit $M_{f} \rightarrow \infty$, 
and the result of the vacuum polarization in the PQED is recovered. 
The inverse of gauge propagator corrected by the vacuum polarization at one loop 
is known as the Schwinger-Dyson formula 
\begin{eqnarray}
D_{\bar{\mu}\bar{\nu}}^{-1}(k_{0},{\bf k})=\Delta_{\bar{\mu}\bar{\nu}}^{-1}(k_{0},{\bf k})-\Pi_{\bar{\mu}\bar{\nu}}(k_{0},{\bf k}) \; .
\end{eqnarray}
The static potential at one loop is defined by
\begin{equation}\label{U1}
U^{(1)}({\bf r})=- e^2\int \frac{d^{2}{\bf k}}{(2\pi)^2} \frac{ e^{i\,{\bf k}\cdot{\bf r}} }{\Delta_{00}^{-1}(k_0=0,{\bf k})-\Pi_{1}^{00}(k_0=0,{\bf k})} \; ,
\end{equation}
that in the case of $(M,M_{f},m) \gg |{\bf k}|$, we obtain 
\begin{eqnarray}\label{U1approx}
U^{(1)}(r) &\simeq& -\frac{e^2}{4\pi}\left[ \, \frac{1-e^{-Mr}}{r}-\frac{\pi\theta}{2} I_{0}(\theta r)+\frac{\pi\theta}{2} L_{0}(\theta r) \, \right] 
\nonumber \\
&&
\hspace{-1.0cm}
+\frac{4\,e^4\,\gamma\,M^2}{3\pi^2[(Mr)^2-4][(\theta r)^2-4]^2}\left(\frac{1}{m}+\frac{1}{M_{f}} \right) \; , \; \; \; \;
\end{eqnarray}
where $\gamma=0.577$ is the Euler-Mascheroni constant. The first line confirms the result (\ref{Urtreelevelresult}), and the second line in (\ref{U1approx}) yields the contribution to a Lamb shift in a planar theory as function of the LW masses $M$, $M_{f}$, and of the $\theta$-parameter. Since we know the condition of $M_{f} \gg m$, 
the contribution of $M_{f}$ to the Lamb shift potential is very small in relation to the electron mass, and we obtain the variation of potential energy  
\begin{equation}
\Delta U = U^{(1)}-U \simeq \frac{4\,e^4\,\gamma\,M^2/m}{3\pi^2[(Mr)^2-4][(\theta r)^2-4]^2} \; . 
\end{equation}
The integral has only exact solution numerically. After the angular integration, it is reduced to $k$-integral
\begin{equation}\label{U1}
U^{(1)}(r)=- e^2 \int_{0}^{\infty} \frac{dk}{2\pi} \, \frac{ k \, J_{0}(kr)}{\Delta_{00}^{-1}(k)+e^2 \, k^2 \, \Pi_{1}(k)/\pi} \; .
\end{equation}
The numerical solution of (\ref{U1}) versus the radial distance $r$ is illustrated in the figure (\ref{Fig2}). 
We choose the values for the masses $m=0.5$ (electron mass) , $M = 100$ (Lee-Wick mass) ,  $M_{f} = 50$ 
(Lee-Wick fermion partner) and the Fermi velocity as $\beta =0.003$ for the graphene that leads to fundamental charge $e=0.016$. 
The colored lines show the $\theta$-values : $\theta=0$ (black dashed line) , $\theta=1.0$ (red line), $\theta=2.0$ (blue line) and $\theta=3.0$ (green line). 
The radiative correction of the vacuum polarization yields a positive contribution to the potential when it is compared 
to figure (\ref{Fig1}).
%
%
%

%
In the case of $|{\bf k}| \gg (M,M_{f},m)$, the integral (\ref{U1}) yields a analytical result
\begin{equation}\label{Urhighk}
U^{(1)}(r)= -\frac{e^2M^2}{8\theta} 
\left[ \, I_{0}\left(\theta\,r\right)-L_{0}\left( \theta \, r \right) \, \right] \; ,
\end{equation}
where $I_{0}(x)$ is a 
modified Bessel function, and $L_{0}(x)$ is a modified Struve function. 
For a reference in special functions, see \cite{Gradstein}. In the limit $r\rightarrow 0$, the static energy is finite at origin
\begin{eqnarray}
U^{(1)}(r \rightarrow 0)
\simeq -\frac{e^2M^2}{8\,\theta} \; .
\end{eqnarray}
%

%
%


%
The $3$-vertex correction at one loop for the $\psi$-fermion is given by
\begin{equation}\label{Vertex}
\Lambda_{\psi}^{\bar{\mu}}(\bar{p}^{\prime},\bar{p}) = \int\frac{d^{3}\bar{k}}{(2\pi)^{3}} \, \Delta_{\bar{\alpha}\bar{\beta}}(\bar{k}) 
V^{\bar{\alpha}} S_{F}^{\psi}(\bar{p}^{\prime}-\bar{k})
V^{\bar{\mu}} S_{F}^{\psi}(\bar{p}-\bar{k}) V^{\bar{\beta}} ,
\end{equation}
where $\Delta_{\bar{\alpha}\bar{\beta}}(\bar{k})$ is the gauge propagator (\ref{propapprox}), $\bar{p}^{\prime}$ and $\bar{p}$ are the external momentum 
associated with the fermion lines, and the transfer momentum for the photon line is $q^{\bar{\mu}}=p^{\prime\bar{\mu}}-p^{\bar{\mu}}$. 
By power counting, this integral has a logarithmic divergence with degree $D=5-2-2-1=0$ in the ultraviolet limit. 
But, since we know from the previous loop integrals, the dimensional regularization is introduced, and after the calculus, the limit to three dimensions 
yields a finite result due to analytic extension of the Gamma functions. 
%

\vspace{0.5cm}
\begin{figure}
\includegraphics[width=\linewidth]{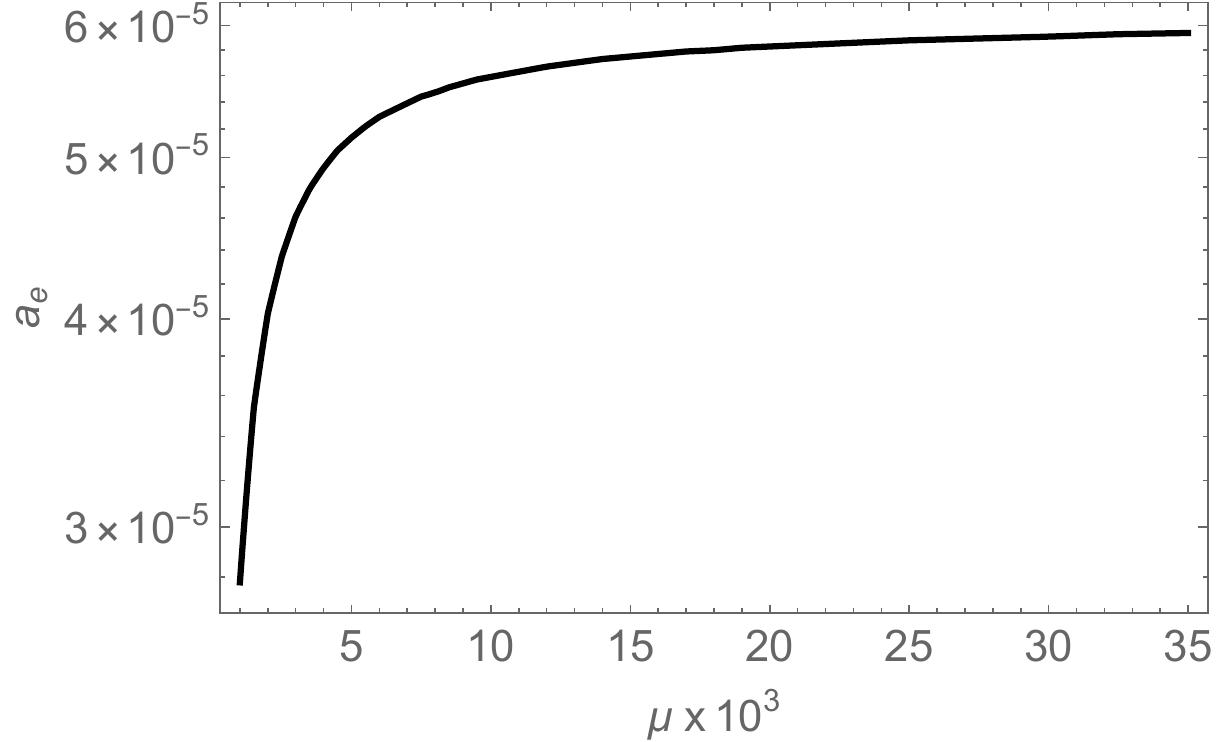}
\caption{The correction to the electron anomalous momentum as function of dimensionless parameter $M/m$ (the Lee-Wick mass over the electron mass). The curve is plotted for $\beta=0.003$ (graphene), with the CS parameter of $\theta=0.1$, and the fine structure constant $\alpha=1/137$. } \label{Fig3}
\end{figure}

It is worth to highlight that the time-like component $(\bar{\mu}=0)$ is not similar to spatial component $(\bar{\mu}=i)$ in the $3$-vertex function due to presence of Fermi velocity that breaks the spatial-time symmetry in the fermion sector. The spatial component $\Lambda_{\psi}^{i}$ in (\ref{Vertex}) that contributes to the electron's $g-2$ factor through the form factor. We write it as $\Lambda_{\psi}^{i}=\Lambda^{i}+\Lambda_{g}^{i}$. The $(g-2)$-contribution is $M^{i}=\overline{u}(\bar{p}) \, \Lambda_{g}^{i} \, u(\bar{p}^{\prime})$, and using the dimensional regularization with the identity (\ref{ID}), the limits of $D \rightarrow 3$ and $q^{2}\rightarrow 0$, we obtain the result 
\begin{eqnarray}
M^{i}=ie\,\beta \, \overline{u}(\bar{p})\left[ \, \frac{1}{2m} \, F_{2}(0) \, \beta \, \sigma^{i\nu} q_{\nu} \, \right]u(\bar{p}^{\prime}) \; ,
\end{eqnarray}
where the form factor $F_{2}(0)$ is given by
\begin{eqnarray}\label{F2}
F_2(0)=\frac{\alpha\,\beta^3}{2\pi} \int_{0}^{1}dx \int_{0}^{1-x} dy \, \Delta^{-1} 
\nonumber \\
\times \, \frac{2+4\beta^2+\beta^2(x+y)}{\sqrt{1-x-y}}
\nonumber \\
\times \left[\frac{1}{x+y}+\frac{x+y}{\mu^2(1-x-y)\Delta-(x+y)^2}   \right] 
\nonumber \\
+\theta \, \frac{\alpha\,\beta^3}{8}\int_{0}^{1}dx\int_{0}^{1-x}dy\,\Delta^{-5/2}\left\{\frac{1}{(x+y)^2}
\right.
\nonumber \\
\left.
+\frac{x+y}{[\,\mu^2(1-x-y)+(x+y)^2\,]^{3/2}} \right\}
\nonumber \\
-\theta\, \frac{\alpha\, \beta^3}{4\pi}\int_{0}^{1} dx \int_{0}^{1-x}dy \int_{0}^{1-x-y}dz\,\Delta^{-5/2}
\nonumber \\
\times \frac{x+y}{\sqrt{ z\,(1-x-y-z) }} \frac{1}{[\, \mu^2\,z+(x+y)^2 \, ]^{3/2}} 
\; ,
\end{eqnarray}
and $\Delta=1+(\beta^2-1)(x+y)$, with $\mu=M/m$. The electron anomalous momentum is defined by $a_e=(g-2)/2=F_{2}(0)$, in which it is 
function of the parameters $\mu$, $\theta$ (CS parameter) and of the Fermi velocity $\beta$. In this result, we have considered $\bar{k}^2 \gg \theta^{2}$, 
where the effects of $\theta$ are very small when compared to LW mass, electron mass, and the momentum integration $\bar{k}$. The numeric solution of (\ref{F2}) is 
illustrated in the figure (\ref{Fig3}) as function of $\mu=M/m$. The curve is drawn for the case of graphene with $\beta=0.003$, 
$\theta=0.1$, and the fine structure constant $\alpha=1/137$. In the limit $\mu \rightarrow \infty$ is equivalent to LW mass goes 
to infinity, that leads to the result $a_{e} \sim 6 \times 10^{-5}$ for the particular case of the PQED with Fermi velocity $\beta=0.003$ 
in the presence of a CS term of $\theta=0.1$.
The term of $\Lambda^{i}$ that contributes to the spatial component of the vertex function is given by 
\begin{widetext}
\begin{eqnarray}\label{Lambdai}
&&
\Lambda^{i}(m) = \frac{e^2}{32\pi} \, \beta \, \gamma^{i} \int^{1}_{0} dx \, \frac{\sqrt{1-x}}{ (1-(1-\beta^2)x)^2 }
\,
\mbox{ln}\left[\frac{ x^2 }{ \mu^2(1-x) + x^2 }\right]
\nonumber \\
&&
-\frac{\theta\,e^2}{32\pi}\,\beta^2\, \gamma^{i}\,\int_{0}^{1} dx \, \frac{1}{ (1-(1-\beta^2)x)^2 }
\left[ 1+\frac{x}{ \sqrt{ \mu^2(1-x)+x^2 }} \right]
\nonumber \\
&&
+\frac{\theta\,e^2m}{8\pi M}\,\beta^2 \, \gamma^{i} \, \int_{0}^{1} dx \, \frac{x}{\sqrt{1-x}} \, \frac{1}{(1-(1-\beta^2)x)^2} 
\left\{ \, K\left[ 1+\frac{x^2}{(1-x)\mu^2} \right] + i \, K\left[\frac{-\,x^2}{(1-x)\mu^2} \right] \, \right\} 
\; .
\end{eqnarray}
\end{widetext}
The result (\ref{Lambdai}) diverges in the limit $M \rightarrow \infty$. 
As consequence of the Ward identities, the $3$-vertex function is finite in $1+2$ dimensions 
in which the Lee-Wick mass is a natural regulator parameter that turn the LWCSPED a finite QFT in the first loop corrections.

\section{The discussion of the unitarity in LWCSPED}
\label{sec6}
Unitarity is one of the fundamentals characteristics of a QFT, that motivates us to prove if the LWCS pseudo-ED has such property. 
The condition of a unitary field theory is associated with the $S$-matrix to be a unitary operator, that is, $S^{\dagger}S=\mathds{1}$. Writing the $S$-operator 
as $S=\mathds{1}+i\,T$, the unitarity condition yields the relation $i\,(T^{\dagger}-T)=T^{\dagger}\,T$, that in terms 
of the initial $(i)$, and final $(f)$ states, the matrix elements are given by
\begin{eqnarray}\label{ImTii}
2\,\mbox{Im}(T_{ii})=\sum_{f}T_{if}^{\dagger}\,T_{fi}=\sum_{f}T_{fi}^{\ast}\,T_{fi} \; ,
\end{eqnarray}   
where $T_{fi}= \langle f |T| i \rangle$. The relation (\ref{ImTii}) is known as the Optical theorem. The elements $T_{ii}$ 
are written in terms of the Feynman Green function as $T_{ii}=(2\pi)^3\,\delta^3(0)\,\Delta_{F}(\bar{x}-\bar{x}^{\prime})$, 
thus, the Optical theorem leads to relation
\begin{eqnarray}\label{RelDeltaF}
\Delta_{F}^{\ast}(\bar{x})-\Delta_{F}(\bar{x})=-i\int d\Phi \, (2\pi)^3\,\delta^{3}(0)
\nonumber \\
\times \int\frac{d^3\bar{x}^{\prime}}{(2\pi)^3} \, \Delta_{F}^{\ast}(\bar{x}^{\prime})
\,\Delta_{F}(\bar{x}-\bar{x}^{\prime}) \; ,
\end{eqnarray}   
in which $d\Phi$ is a phase factor for the dimensional analysis of this relation leads to identity. It defines 
the characteristic time $({\cal T})$ of the system by the phase factor
\begin{eqnarray}
\int d\Phi \, (2\pi)^3\,\delta^{3}(0)={\cal T}^{-1} \; .
\end{eqnarray} 
In the momentum space, the relation (\ref{RelDeltaF}) for a gauge field theory is given by
\begin{eqnarray}\label{GFrelgauge}
\Delta_{\bar{\mu}\bar{\nu}}^{\ast}(\bar{k})-\Delta_{\bar{\mu}\bar{\nu}}(\bar{k})=-i \, {\cal T}^{-1} \, \Delta_{\bar{\mu}\bar{\alpha}}^{\ast}(\bar{k})\,\Delta^{\bar{\alpha}}_{\;\;\,\bar{\nu}}(\bar{k}) \; . \;\;\;\;
\end{eqnarray}
After this brief summary, we firstly investigate the unitarity condition at tree level 
for the LWCS pseudo-ED ruled by the gauge propagator (\ref{propapprox}). This propagator has a form
\begin{eqnarray}\label{propgaugeS}
\Delta_{\bar{\mu}\bar{\nu}}(\bar{k})=A(\bar{k})\,\theta_{\bar{\mu}\bar{\nu}}+B(\bar{k})\,S_{\bar{\mu}\bar{\nu}} \; ,
\end{eqnarray}
where 
\begin{subequations}
\begin{eqnarray}
A(\bar{k}) &=& \frac{-i\,\bar{k}^2}{2(\bar{k}^2-\theta^{2}+i\epsilon)}
\nonumber \\
&&
\hspace{-0.4cm}
\times\left[ \frac{1}{\sqrt{-\bar{k}^2-i\epsilon}}-\frac{1}{\sqrt{-\bar{k}^2+M^2-i\epsilon}} \right] \; ,
\label{Ak}
\;\;\;\;
\\
B(\bar{k}) &=& \theta\, (\bar{k}^2)^{-1} \, A(\bar{k}) \; ,
\label{Bk}
\end{eqnarray}
\end{subequations}  
and we have added the Feyman prescriptions $\theta^2 \rightarrow \theta^2-i\epsilon$ and $ M^2 \rightarrow M^{2}-i\epsilon$.
Substituting (\ref{propgaugeS}) in (\ref{GFrelgauge}), we obtain the relations  
\begin{subequations}
\begin{eqnarray}
2\,\Im[A(\bar{k})] &=& -{\cal T}^{-1} \left[ A^{\ast}(\bar{k})A(\bar{k})+\bar{k}^2\,B^{\ast}(\bar{k})B(\bar{k}) \right] , \;\;\;\;
\label{ImA}
\\
2\,\Im[B(\bar{k})] &=& -{\cal T}^{-1} \left[ A^{\ast}(\bar{k})B(\bar{k})+B^{\ast}(\bar{k})A(\bar{k}) \right] \; ,
\label{ImB}
\end{eqnarray}
\end{subequations}
in which $\Im$ are the imaginary parts of $A$ and $B$. We can write (\ref{Ak}) as
\begin{eqnarray}
A(\bar{k}) &=& \frac{-i\,\bar{k}^2}{2} \, \frac{\bar{k}^2-\theta^{2}-i\epsilon}{(\bar{k}^2-\theta^{2})^2+\epsilon^2}
\nonumber \\
&&
\hspace{-0.4cm}
\times\left[ \frac{\sqrt{-\bar{k}^2+i\epsilon}}{\sqrt{(\bar{k}^2)^2+\epsilon^2}}-\frac{\sqrt{-\bar{k}^2+M^2+i\epsilon}}{\sqrt{(\bar{k}^2+M^2)^2+\epsilon^2}} \right] \; ,
\label{Akconju}
\end{eqnarray}  
where the complex variables in polar form are
\begin{subequations}
\begin{eqnarray}
\sqrt{-\bar{k}^2+i\epsilon} &=& \sqrt{\rho_{k}}\,e^{i\alpha/2}
\; ,
\label{kcomplex}
\\
\sqrt{-\bar{k}^2+M^{2}+i\epsilon} &=& \sqrt{\rho_{M}}\,e^{i\delta/2}
\; ,
\label{kMcomplex}
\\
\bar{k}^2-\theta^2-i\epsilon &=& \rho_{\theta}\, e^{i\gamma}
\label{kthetacomplex}
\; ,
\end{eqnarray}
\end{subequations}
with $\rho_{k}=\sqrt{(\bar{k}^2)^2+\epsilon^2}$, $\rho_{M}=\sqrt{(\bar{k}^2-M^2)^2+\epsilon^2}$, $\rho_{\theta}=\sqrt{(\bar{k}^2-\theta^{2})^2+\epsilon^2}$, 
and the phases are defined by $\alpha=\sin^{-1}(\epsilon/\rho_{k})$, $\delta=\sin^{-1}(\epsilon/\rho_{M})$ and $\gamma=\sin^{-1}(-\epsilon/\rho_{\theta})$. 
Using these definitions, the relation (\ref{ImA}) is
\begin{eqnarray}\label{ImAparameters}
\frac{1}{\sqrt{\rho_k}}\,\cos\left( \frac{\gamma+\alpha}{2}\right)
-\frac{1}{\sqrt{\rho_M}}\,\cos\left( \frac{\gamma+\delta}{2}\right)=
\nonumber \\
={\cal T}^{-1} \, \frac{(\bar{k}^2+\theta^2)}{4\,\rho_{\theta}}\left[\frac{1}{\rho_{k}}+\frac{1}{\rho_{M}}-\frac{2}{\sqrt{\rho_k\rho_M}}\,\cos\left(\frac{\alpha-\delta}{2} \right) \right] \, .
\nonumber \\
\end{eqnarray}
Squaring (\ref{ImAparameters}) and multiplying it by $\epsilon$, the identities for small $\epsilon$-parameter  
\begin{subequations}
\begin{eqnarray}
\pi\,\delta(\bar{k}^2) &=& \frac{\epsilon}{(\bar{k}^2)^2+\epsilon^2} \; ,
\label{deltak}
\\
\pi\,\delta(\bar{k}^2-M^2) &=& \frac{\epsilon}{(\bar{k}^2-M^2)^2+\epsilon^2} \; ,
\label{deltakM}
\\
\pi\,\delta(\bar{k}^2-\theta^2) &=& \frac{\epsilon}{(\bar{k}^2-\theta^2)^2+\epsilon^2} \; ,
\label{deltaktheta}
\end{eqnarray}
\end{subequations}
yield the relation
\begin{eqnarray}\label{EqSimp}
\epsilon \cos^2\left(\frac{\gamma_0}{2}+\frac{\pi}{4}\right)+\epsilon \cos^2\left(\frac{\gamma_M}{2}+\frac{\pi}{4}\right)-
\nonumber \\
-\frac{2\,\epsilon^{3/2}}{(M^4+\epsilon^2)^{1/4}}\cos\left(\frac{\gamma_0}{2}+\frac{\pi}{4}\right)\cos\left( \frac{\gamma_0+\delta_{0}}{2} \right)=
\nonumber \\
={\cal T}^{-2}\,\frac{\theta^4}{4}\left[ \frac{1}{\sqrt{\theta^4+\epsilon^2}}+\frac{1}{\sqrt{(M^2-\theta^2)^2+\epsilon^2}}
\right.
\nonumber \\
\left.
-\frac{2 \, \cos\left( \frac{\gamma_0-\gamma_{M}}{2} \right)}{(\theta^4+\epsilon^2)^{1/4} ((M^2-\theta^2)^2+\epsilon^2)^{1/4} }
\right]^2 ,
\end{eqnarray}
where $\gamma_{0}$ and $\delta_{0}$ are the $\gamma$- and $\delta$-phases evaluated at $\bar{k}^2=0$, respectively, 
and $\gamma_{M}$ is the $\gamma$-phase evaluated at $\bar{k}^2=M^2$. Namely, we write $\gamma_{0}$, $\delta_{0}$ and $\gamma_{M}$ below :
\begin{subequations}
\begin{eqnarray}
\gamma_{0} &=& \sin^{-1}\left[\frac{-\epsilon}{\sqrt{\theta^4+\epsilon^2}}\right] \; ,
\\
\delta_{0} &=& \sin^{-1}\left[\frac{\epsilon}{\sqrt{M^4+\epsilon^2}}\right] \; ,
\\
\gamma_{M} &=& \sin^{-1}\left[\frac{-\epsilon}{\sqrt{(M^2-\theta^2)^2+\epsilon^2}}\right] \; .
\end{eqnarray}
\end{subequations}
For the condition of $M \gg \theta \gg \sqrt{\epsilon}$, the solution of (\ref{EqSimp}) for the characteristic time $({\cal T})$ is given by
\begin{eqnarray}\label{SolT}
{\cal T}^{-1}\simeq 2\,\sqrt{\epsilon}\left(1+\frac{2\theta}{M}\right) \; .
\end{eqnarray}
Notice that in the limit $\theta \rightarrow 0$, or in the case of $M \rightarrow \infty$, the result of the usual PED is recovered. 
Now we need to check that the second relation (\ref{ImB}) also provides a solution for the characteristic time. Using the 
parameterizations of the complex variables (\ref{kcomplex})-(\ref{kthetacomplex}), with the previous manipulations, the relation (\ref{ImB}) leads to same 
expression (\ref{EqSimp}). Thereby, the equation (\ref{SolT}) is the solution of (\ref{ImA}) and (\ref{ImB}), and we can confirm that the LWCSPED 
is unitary gauge field theory at tree level.     
\section{Conclusions}
\label{sec7}
The Lee-Wick (LW) pseudo-electrodynamics added to a non-local Chern-Simons (CS) topological term in $1+2$ dimensions is investigated in this paper.
We obtain the gauge propagator and the correspondent poles in which two massive poles emerge : the heavy LW mass, and a light mass associated with the CS parameter.
The photon dispersion relation is also one of the poles as consequence of the gauge invariance. These massive degree of freedom are interpreted as the masses in the gauge sector without the need of introducing scalar fields via spontaneous symmetry breaking mechanism. Thereby, this is the advantage of the LWCS approach.
Using the gauge propagator with static sources, the potential energy of interaction between two electrons is calculated 
as function of the radial distance on the plane that separates these two charges, and also in terms of LW mass, and of CS mass parameter.
We use the gauge propagator to obtain the retarded Green function via Fourier transformation, and consequently, the result show that the theory respects 
the causality principle. The sector of fermions is so included using the Lee-Wick approach, where a massive degree of freedom is interpreted as a heavy fermion, 
beyond the usual light massive electron. These fermions are coupled to the LWCS gauge field that preserve the $U(1)$-local gauge invariance. The theory is so called 
Lee-Wick-Chern-Simons pseudo-quantum electrodynamics (LWCSPQED). We construct the generating functional and the effective action approach 
of this theory that leads to the Ward identities. Although the fermion sector introduces new physical degree of freedom through the heavy 
LW fermion, it contribution at one loop approximation is very small when compared to electron mass, the LW mass, and the CS parameter.   
Posteriorly, we calculate the quantum corrections at one loop to the electron's propagator 
(electron's self-energy), to the gauge propagator (vacuum polarization), and to the $3$-vertex of the electron-positron pair interacting with 
the LWCS gauge field. We use the dimensional regularization and the technics of QFT to evaluate the loop integrals. 
Taking the limit to the three dimensions, the integrals are finite by the analytic extension of the Gamma functions, and also by the 
presence of the LW mass $(M)$ that works as a natural regulator parameter. In the limit of $M \rightarrow \infty$, the ultraviolet divergences are recovered at one loop.  
Thus, the self-energy provides a correction to the electron mass that is finite in $1+2$ dimensions in the LWCSPQED. A finite result also is obtained for the vacuum 
polarization in which the potential energy of interaction between two static electrons is calculated numerically in the presence of the quantum corrections at one loop.
We obtain energy potential for the case of small the external photon momentum in relation to masses, that yields a contribution to the Lamb shift, 
eq. (\ref{U1approx}). For higher external momentum, the energy potential is ruled by modified Struve and Bessel functions, as showed in the eq. (\ref{Urhighk}).
The correction to the $3$-vertex is also calculated via technics of QFT in which we extract the form factor $F_{2}(0)$ that contributes to the electron's anomalous momentum 
in the LWCSPED, {\it i. e.}, $a_e=(g-2)/2=F_{2}(0)$, as show the result (\ref{F2}). Since $F_{2}$ is function of the LW mass $(M)$ over the electron mass $(m)$, 
the numeric result of $F_2$ versus $\mu=M/m$ is showed in the fig. (\ref{Fig3}). The limit result for a very large LW mass leads to $a_e \sim 6 \times 10^{-5}$, that it is 
exactly the result of the usual PED with CS term. The complete result for the $3$-vertex also is finite in $1+2$ dimensions, and diverges only in the limit of $M \rightarrow \infty$, and we recover the results of the Maxwell-Chern-Simons PQED. 
Other important characteristic of the LWCSPED as a QFT is the unitarity, that it is investigated in the last section of the paper. Using the Optical theorem, we show that the theory is unitary at tree level, even in the presence of the topological and non-local CS term. 
As perspective of a forthcoming project, the theory opens the possibility of investigation for a Lee-Wick scalar sector in $1+2$ dimensions 
coupled to the LWCS gauge field, in which the quantum effects at one loop may yield contributions to the effective potential.         
\end{document}